\documentstyle[prb,multicol,aps,epsf,graphicx]{revtex}

\newcommand \bi {\bibitem}
\newcommand \bc {\begin{center}}
\newcommand \ec {\end{center}}
\newcommand \ee {\end{equation}}
\newcommand \be {\begin{equation}}
\newcommand \beq {\begin{eqnarray}}
\newcommand \eeq {\end{eqnarray}}
\newcommand \bmu {\begin{multline}}
\newcommand \emu {\end{multline}}
\newcommand \e {\epsilon}
\newcommand \eps {\epsilon}

\title{Validity of the zero-thermodynamic law in off-equilibrium
coupled harmonic oscillators} 
\author{A.Garriga and F.Ritort}
\address{Department of Physics, Faculty of Physics, University of
Barcelona\\ Diagonal 647, 08028 Barcelona (Spain) }

\begin{document}
\maketitle

\begin{abstract}
In order to describe the thermodynamics of the glassy systems it has
been recently introduced an extra parameter, the effective
temperature which generalizes the fluctuation-dissipation theorem
(FDT) to off-equilibrium systems and supposedly describes thermal
fluctuations around the aging state. Using this concept we investigate the applicability of a zeroth thermodynamic law for non-equilibrium systems. In particular we study two coupled systems of harmonic oscillators with Monte Carlo dynamics. We analyze in detail two types of dynamics: 1) sequential dynamics where
the coupling between the subsystems comes only from the Hamiltonian
and 2) parallel dynamics where there is a further coupling between the
subsystems arising from the dynamics. We show that the coupling
described in the first case is not enough to make asymptotically the
effective temperatures of the two interacting subsystems equalize, the
reason being the too small thermal conductivity between them in the aging
state. This explains why different interacting degrees of freedom in
structural glasses may stay at different effective temperatures without
never mutually thermalizing.
\end{abstract}

\section{INTRODUCTION}
The dynamics of glassy systems has been a subject of intensive
research \cite{SIT}. Despite the fact that glassy systems are 
off-equilibrium systems, some regularities that allow the rationalization
of the problem have been found.  One of the most striking regularities
is the presence of aging. This means that the correlation and response
functions are not only functions of time-differences but also of the
time elapsed since the system was prepared \cite{BOU}. Thus,
qualitatively, the longer one waits in the low temperature phase,
the smaller the response to an external field will be. A salient feature of
systems in equilibrium is the fact that the linear response functions
and the equilibrium fluctuations are related by the well known
fluctuation-dissipation theorem (FDT) \cite{KUB}. This relation does
not hold for off-equilibrium systems. Several studies of
spin-glass mean-field models have shown that a generalization of the
fluctuation-dissipation theorem is possible through the definition of
the ``fluctuation-dissipation ratio'' (FDR)\cite{CUG,FRA}:

\be
X(t,s) = \frac{TG(t,s)}{\frac{\partial C(t,s)}{\partial s}}  (t\geq s)
\label{eqX},\ee

\noindent
which is equal to 1 in equilibrium. It turns out that the behavior of
the quantity $X(t,s)$ is non trivial in the limit $t,s\to\infty$. If
the lowest time $s$ is sent to infinity the quantity $X(t,s)$ becomes
a non-trivial function of the autocorrelation $C(t,s)$. This a strong
statement which has been proved to hold in the framework of mean-field
spin glasses \cite{CUG,FRA}. Moreover, it has been recently recognized that
the quantity $X$ is generally related to the Parisi order parameter
$P(q)$ which appears in equilibrium studies of spin-glasses providing a
natural link between the static and dynamical properties \cite{FMPP}.

What is the physical interpretation of $X$? According to relation
(\ref{eqX}) the fluctuation-dissipation relation would be satisfied if
the temperature into the right hand side of (\ref{eqX}) were
$T/X(t,s)$.  This last ratio receives the name of effective
temperature and it has been shown \cite{KUR} that it has some of the
good properties of a macroscopic temperature. In fact a proper thermometer coupled to the slow degrees of freedom
can measure it. The value of $ T_{\rm eff}(t,s)= T/X(t,s)$ would then be
different (and higher) than that of the thermal bath. The question
about the convenience of this temperature to describe the
non-equilibrium behavior has been a subject of controversy in the
last years \cite{NIEW2}. While there are some evidences (not only
theoretical but also experimental \cite{EXP,SIT}) that the
violation of FDT gives a good temperature in the thermodynamic sense,
it is unclear what properties of standard (i.e. equilibrium)
temperatures are common to the non-equilibrium ones.

The motivation of this paper is to answer to the following question:
How effective temperatures equalize when two systems out of
equilibrium are put in contact? In other words, does there exist a
zeroth law for non-equilibrium systems? Let us imagine about a
vitrified piece of silica quenched to the room temperature. Because
the glass is off-equilibrium its effective temperature is higher than
room temperature. But, if we touch the piece of glass it is not hotter
than the room temperature. We must conclude that some degrees of
freedom within the piece of silica are thermalized to the room
temperature while other remain non-thermalized and still
hotter. Touching the piece of silica we feel the fast modes, not the
slow ones. This poses the question, how is that possible that different
interacting degrees of freedom have not reached thermal equilibrium
for sufficient long times? Despite of some considerations present in
the literature \cite{KUR,CUKU} there are no clear answers to this
question. We believe that some of them may require a more deep
understanding through a detailed analysis of an illustrative example
as a previous stage to offer more simple and generic considerations.
It is our purpose here to follow this route trying to give a
general answer to this question by deriving exact results in the
framework of a solvable model.

The model is a set of harmonic oscillators evolving by Monte Carlo
dynamics introduced in \cite{BPR} (hereafter referred as BPR
model). The importance of this model relies on the fact that it is
exactly solvable and shows one of the main features of glasses, namely
aging in correlation and response functions.  Our interest will be in
considering two coupled sets of harmonic oscillators. Thus, we can see
how the main observables are affected by the coupling, in particular
how the effective temperature evolves for the two sets of interacting
degrees of freedom (represented by the two different sets of harmonic
oscillators). The interaction may then appear through the Hamiltonian
or through the Monte Carlo dynamics itself.  We will discover that the
effective temperature for the two sets of oscillators depends on how
the coupling is done, and we will understand why in vitreous systems different
degrees of freedom may stay at different temperatures without
thermalising at very long times. The central idea is that interacting
non-equilibrium systems each one with very different effective
temperatures may not equalize because the conductivity in the aging
state can be extremely small. In this sense the utility of the
extension of the zeroth thermodynamic law to the non-equilibrium
aging state is questioned due to the smallness of the non-equilibrium
conductivities.

The paper is organized as follows. Section II describes the main aspects
as well as the interest of the model. Section III describes the two
classes of couplings we have considered. Section IV analyzes the case in
which the main coupling is ruled by the Monte Carlo dynamics. Section V
describes the case where coupling appears only in the Hamiltonian. Sections IV and V show how to solve the dynamics of the system. The reader who is not interested in technical issues can skip them. Section VI discusses the results and the physical consequences of our work. The last section presents the conclusions. Three appendices are devoted to
some other technical issues.
 
\section{A SIMPLE AND SOLVABLE MODEL OF GLASS}

As a simple model of glass we will consider a system of uncoupled
harmonic oscillators evolving with Monte Carlo dynamics.  The
Hamiltonian is:
\begin{equation}
H = \frac{1}{2} K \sum_{i=1}^N x_i^2~~~~~.
\end{equation}
This model was introduced in \cite{BPR} and was also reviewed in \cite{TEO,FEL}.\\
The low-temperature Monte Carlo dynamics of an ensemble of linear
harmonic oscillators shows typical non-equilibrium features of glassy
systems like aging in the correlation and response functions. The
interest of this model is that the slow dynamics at low temperatures is
a consequence of the entropy barriers generated by the low acceptance
rate.  The simplicity of this model makes it exactly solvable yielding a
lot of results about the non-equilibrium behavior.

The Monte Carlo move consists on the following: the $ x_i $ are moved to
$ x_i + r_i/\sqrt{N} $ where $ r_i $ are random variables Gaussian
distributed with zero average and variance $ \Delta^2. $ The move is
accepted according to the transition probability $W(\Delta E) $ which
satisfies detailed balance: $ W(\Delta E) = W(-\Delta E)\exp(-\beta
\Delta E ), $ where $ \Delta E $ is the change in the Hamiltonian. In Appendix A we show the computation of the correlation and response functions. Here we only quote the main results,

\begin{enumerate}

\item{\em Slow decay of the energy.} The evolution equation for the
energy is Markovian. This simplicity allows for an asymptotic large-time
expansion showing that the energy decays logarithmically $E(t)\sim
1/\log(t)$ and the acceptance ratio decays faster $A(t)\sim
1/(t\,\log(t))$.

\item{\em Aging in correlations and responses.} The correlation function
$C(t,s)$ is defined by:
\begin{equation}
 C(t,s) =\frac{1}{N} \sum_{i=1}^N x_i(t)  x_i(s)~~~~~. 
\end{equation}

The response function is calculated by applying an external field to the
system. Then, the response function is the variation of the
magnetization of the system when the
field is applied:

\begin{equation}
G(t,s) = \left( \frac{\delta M(t)}{\delta h(s)} \right)_{h=0}~~~~~t>s~~~~,
\end{equation} 
with the magnetization given by,
\begin{equation}
M(t)=\frac{1}{N}\sum_{i=1}^N x_i(t)~~~~~.
\label{eqM}
\end{equation}

Details on how to solve correlations and responses are given in
Appendix A. The final results are equations
(\ref{ApC},\ref{ApG}). Both correlation and responses show dominant
$s/t$ scaling with logarithmic corrections. The asymptotic scaling
behavior is given by,

\be
C(t,s)=C(s,s)\frac{L(s)}{L(t)}~~~~~~,~~~~~~~
G(t,s)=G(s,s)\frac{L(s)}{L(t)}\Theta(t-s)~~~,\label{eqG}
\ee

with $C(s,s)=\frac{2E(s)}{K}, G(s,s)=\frac{f(s)}{K}$ where the
expression $f(t)$ is given in equation (\ref{ft}) and $L(t)\sim
t(\log^2(t))$ \cite{THEO2}.  The slow decay of the response
function shows the presence of long-term memory which manifests as aging
in the integrated response function \cite{BPR}. 

\item{\em The effective temperature.}
As said in the introduction, the effective temperature is defined in
terms of the FDR eq.(\ref{eqX}):

\begin{equation}
T_{\rm eff}(t,s)=\frac{\frac{\partial C(t,s)}{\partial s}}{G(t,s)}\label{eq_Teff}~~~~~.
\end{equation}


In equilibrium $E(s)=T/2$ and we recover the expected result $T_{\rm
eff}=T$. Interestingly (\ref{eq_Teff}) yields a result for $T_{\rm
eff}$ which only depends on the smallest time $s$. The unique dependence
of the effective temperature on the lowest time $s$ is generally
believed to be satisfied in the asymptotic large $s$ limit for generic
structural glasses and spin-glass models with a one step of replica
symmetry breaking. This expectation holds here for all times. At zero temperature when
slow motion sets in, the system never reaches the ground state and ages
forever.  In this regime the effective temperature verifies in the
long-time limit (i.e $s \longrightarrow \infty$ ):

\begin{equation}
T_{\rm eff}(s) = 2E(s) + \frac{2}{f(s)} \frac{\partial E(s)}{\partial s} \longrightarrow 2E(s),
\label{eq_Teff3}
\end{equation}

This gives a thermodynamic relationship between the effective
temperature and the dynamical energy in the off-equilibrium regime
showing how the equipartition theorem can be extended to the glassy
regime. The effective temperature measures how a quasi-stationary or
adiabatic hypothesis is exact for the present model suggesting that some
features of equilibrium thermodynamics may be applied to the aging
regime.

\end{enumerate}

\section{TWO COUPLED SYSTEMS}

Now we consider the case in which we couple two systems of harmonic
oscillators. In this case it is possible to compute analytically
how one system affects the other without loosing the benefit of
evaluating Gaussian integrals.  The Hamiltonian we
have to deal with is:

\begin{equation} H = \frac{K_{1}}{2} \sum_{i=1}^N x_i^2 + \frac{K_2}{2} \sum_{i=1}^N y_i^2 - \frac{\epsilon}{N} \sum_{i=1}^N x_iy_i, \label{eqH}\end{equation}
where we take $K_1K_2>\eps^2$, otherwise the system has no bounded
ground state. We define the following extensive quantities (per
oscillator):

\be E_1 = \frac{K_1}{2N} \sum_{i=1}^N x_i^2~~~~~,~~~~~E_2
= \frac{K_2}{2N} \sum_{i=1}^N y_i^2~~~~~,~~~~~Q =
\frac{1}{N}\sum_{i=1}^N x_i y_i \label{q} \ee
where $ E_1 $ and $ E_2 $ are the energy of the bare systems while $Q$ is the
overlap between them.  In this case we also consider Monte-Carlo
dynamics, where the transition probability is performed by the Metropolis
algorithm which satisfies detailed balance. The random changes in the
degrees of freedom $x_i,y_i$ are defined in the same way we have
explained in the previous section for
the case of a single system. But there are different ways to implement the
dynamics in the model depending on the updating procedure of the
variables $x_i,y_i$. Here we have analyzed two important and different
procedures which yield quite different results:

\begin{enumerate}
\item{\em Uncoupled or sequential dynamics.} In this case the two sets of variables $x$
and $y$ are sequentially updated. First the $x_i$ variables are updated
and the move is accepted according to the total change of energy $\Delta
E= \Delta E_1-\eps\Delta Q$. Next, the variables $y_i$ are changed and
the move accepted according to the energy change $\Delta E= \Delta
E_2-\eps\Delta Q$. This procedure is then iterated. In this
case, the dynamics does not affect simultaneously the two sets of
variables but each set is updated independently from the other. The only
coupling between the two sets of oscillators comes from the explicit
coupling term $\eps Q$ in the Hamiltonian. Note that for $\eps=0$ the
dynamics becomes trivial because the dynamical evolutions are that of two
independent sets of harmonic oscillators everything reducing to the
original model described in section II.

\item{\em Coupled or parallel dynamics.} In this first case the
$x_i,y_i$ variables are updated in parallel according to the rule
$x_i\to x_i+r_i/\sqrt{N}$, $y_i\to y_i+s_i/\sqrt{N}$. The transition
probability for that move $W(\Delta E)$ is determined by the change in
the total energy $\Delta E=\Delta E_1+\Delta E_2-\eps \Delta Q$
introducing, on top of the explicit coupling term $\eps Q$ in the
Hamiltonian, an additional coupling between the whole set of
oscillators through the parallel updating dynamics. Contrarily to the
uncoupled case, the $\eps=0$ case is interesting by itself because it
shows how this kind of dynamical coupling strongly influences the
glassy behavior. In fact, in the limiting case $\eps=0$, there will be
some changes which make the energy of one of the two systems increase,
this change being accepted because the total energy will
decrease. Because of that, despite of the fact that there is no direct
coupling in the Hamiltonian the dynamics turns out to be strongly coupled.

\end{enumerate}

\noindent
In what follows we describe the main set of quantities we are
interested in. The solution of the dynamical equations for the coupled
and uncoupled cases is very similar. The Appendix B shows in detail
the derivation of the dynamical solution for the uncoupled case.

\subsection{Correlation, overlaps and responses}

On top of the time evolution of one-time quantities our interest will
also focus on the behavior of two-times quantities such as
correlations and responses. These quantities will refer to three
classes of systems: the set of oscillators described by the $x$
variables, the set of oscillators described by the $y$ variables and
the whole set of $x$ and $y$ variables. In the rest of the paper, as a
rule, the subindex 1 will refer to quantities describing the set $x$
of oscillators, the subindex 2 will refer to quantities describing the
set $y$ of oscillators and the subindex $T$ will refer to quantities
describing the whole set of oscillators $x$ plus $y$. The main set of
correlation and response functions we are interested in are:

\begin{itemize}
\item{\em Correlations.} The correlation function for the sets $x$ and
$y$,

\be C_1(t,s) =\frac{1}{N} \sum_{i=1}^N x_i(t)
 x_i(s)~~~~~,~~~~~C_2(t,s) =\frac{1}{N} \sum_{i=1}^N y_i(t) y_i(s)~~~~~,
\label{a}
\ee                     
as well as the global correlation $C_T(t,s)=\frac{1}{2}(C_1(t,s)+C_2(t,s))$. 

\item{\em Overlaps.} These are cross-correlations involving
different sets of variables:

\be Q_1(t,s) =\frac{1}{N} \sum_{i=1}^N y_i(t)
 x_i(s)~~~~~,~~~~~Q_2(t,s) =\frac{1}{N} \sum_{i=1}^N x_i(t) y_i(s)~~~~~,
\label{b}
\ee
with $Q_1(t,s)=Q_2(s,t)$. As we will see later, it is useful to 
define these two functions $Q_1,Q_2$ which essentially are the same 
overlap function but acting on different time sectors. 

\item{\em Response functions.}
The response function for the sets $x$ and $y$ are defined in the
following way. Define the magnetizations for the two sets of
oscillators $x$ and $y$,

\be
M_1=\frac{1}{N}\sum_{i=1}^N x_i~~~~~,~~~~~M_2=\frac{1}{N}\sum_{i=1}^N y_i~~~~~.
\end{equation}

Consider also two external fields $h_1$ and $h_2$ conjugated
respectively to $M_1$ and $M_2$,

\begin{equation}
 H = \frac{K_{1}}{2} \sum_{i=1}^N x_i^2 + \frac{K_2}{2} \sum_{i=1}^N y_i^2 -
 \sum_i(h_1x_i + h_2y_i) - \e \sum_i x_iy_i~~~~~.
\end{equation} 

We define four types of response functions $G_{1,2},G'_{1,2}$. The
$G_1(t,s),G_2(t,s)$ functions measure the change in the 
magnetization $M_1(t)$, $M_2(t)$ induced by their respective conjugated field $h_1,h_2$ 
applied at time $s$. These are defined by

\begin{equation}
G_i(t,s) = \left( \frac{\delta M_i(t)}{\delta h_i(s)} \right)_{h_i=0}\label{eqGi}~~~~~,
\end{equation} 
where the index $ i=1,2 $ represents each one of the
systems. Apart from these two response functions we may define the global
response function $G_T(t,s)$ as the change in the global magnetization
$M_T=\frac{1}{2}(M_1+M_2)$ induced by a field conjugate to the total magnetization,

\begin{equation}
G_T(t,s) = \left( \frac{\delta M_T(t)}{\delta h(s)}
\right)_{h=0}=\frac{1}{2}\Bigl( G_1(t,s)+G_2(t,s)\Bigr)
\end{equation} 
 
The primed response functions $G_1'(t,s),G_2'(t,s)$ functions measure
the change in the magnetization in each set of oscillators $M_1(t)$,
$M_2(t)$ induced by a conjugated field (respectively $h_2,h_1$)
applied on the other set of oscillators at a time $s$:

\begin{equation}
G_i'(t,s) = \left( \frac{\delta M_i(t)}{\delta h_j(s)} \right)_{h_j=0}
\end{equation}

where the indices $i,j=1,2$ are different $i\ne j$. In the absence of
a coupling term $\eps Q$ in the Hamiltonian (\ref{eqH}) the two response
functions $G'_{1,2}$ vanish but for $\eps\ne 0$  they enter into the
solution of the dynamical equations.

\item{\em Effective temperatures.} From the correlation and response
functions we may define three effective temperatures:  $T_{\rm eff}^1$ for the system
1, $T_{\rm eff}^2$ for system 2 and $T_{\rm eff}^T$ for the
global system. These are defined as follows,

\be T_{\rm eff}^1 = \left( \frac{\frac{\partial C_1(t,s)}{\partial
s}}{G_1 (t,s)} \right)~~~~~,~~~~~T_{\rm eff}^2 = \left(
\frac{\frac{\partial C_2(t,s)}{\partial s}}{G_2 (t,s)}
\right)~~~~~,~~~~~T_{\rm eff}^T = \left( \frac{\frac{\partial
C_T(t,s)}{\partial s}}{G_T (t,s)} \right)~~~~~.
\label{T1}
\ee

We will analyze in detail the three effective temperatures for the
coupled and the uncoupled cases. From them we will learn whether
the systems equalize their temperatures and how they do.

\end{itemize}

\subsection{Equilibrium regime}

Here we present the results for the statics for the general
model (\ref{eqH}). The equilibrium solution is the  stationary state of
the dynamics coinciding for both coupled and uncoupled dynamics. 
The results for the one-time quantities can be simply evaluated from the
partition function,

\be
{\cal Z}=\int_{-\infty}^{\infty}dx\,dy\,\exp(-\beta H),
\label{eqZ}
\ee
which involves simple Gaussian integration. By performing the
appropriate partial derivatives we calculate the different
thermodynamic quantities:

\be E_1^{\rm eq}=E_2^{\rm eq}=
\frac{K_1K_2T}{2(K_1K_2-\epsilon^2)}~~~~~,~~~~~ Q^{\rm
eq}=\frac{\epsilon I}{J}= \frac{2\epsilon E_1^{\rm
eq}}{K_1K_2}~~~~~,\label{eqQ} \ee

\noindent
where the parameters $I,J$ are defined by,

\be I = \frac{E_1}{K_1} \Delta_2^2 +\frac{E_2}{K_2}
\Delta_1^2~~~~~,~~~~~J=\frac{K_1\Delta_1^2}{2}
+\frac{K_2\Delta_2^2}{2} \label{eqI}~~~~~,  \ee

\noindent
where the total energy $E=E_1+E_2-\eps Q$ is given by the equipartition
relation $E=T$. Note that $K_1K_2-\eps^2>0$ in order for $E_1^{\rm eq},
E_2^{\rm eq}$ to be positive.

The equilibrium correlations $C$, overlaps $Q$ and responses $G,G'$
only depend on the time differences. While the precise form of these
functions depends on the particular type of dynamics, the magnetic
susceptibilities do not. These are given by:

\be
\chi_1 = \int_0^\infty G_1(t) dt= \frac{K_2}{K_1K_2-\epsilon^2},~~
\chi_2 = \int_0^\infty G_2(t) dt = \frac{K_1}{K_1K_2-\epsilon^2},~~
\chi_T = \int_0^\infty G_T(t) dt = \frac{1}{2}(\chi_1+\chi_2),
\label{ji}
\ee

\noindent
and are temperature independent as expected for oscillator
systems. Nonetheless, in equilibrium the three effective temperatures
(\ref{T1}) coincide with the bath temperature $T$.

\section{THE DYNAMICALLY UNCOUPLED (OR SEQUENTIAL) CASE}
In this section we solve the dynamics of the thermodynamic relevant quantities for the case in which the two subsystems of oscillators are dynamically uncoupled. As explained in the previous section, in this case we make a sequential
dynamics avoiding direct dynamical coupling effects coming from the Monte
Carlo dynamics. The derivation of the dynamical equations is explained
in the Appendix B. The equations for the energies and overlap
(\ref{q}) are written down in (\ref{eqApBE1},\ref{eqApBE2},\ref{eqApBE3}),

\beq
\frac{\partial E_1}{\partial t} &=& - ( 2E_1 - \epsilon Q) f_{R_1}(t) + \frac{1}{2} \left( \frac{ f_{R_1}(t)}{\beta} + \frac{K_1\Delta_1^2}{2} ercf(\alpha_1) \right)\label{eqIIIE1}\\
\frac{\partial E_2}{\partial t} &=& - ( 2E_2 - \epsilon Q) f_{R_1}(t) + \frac{1}{2} \left( \frac{ f_{R_2}(t)}{\beta} + \frac{K_2\Delta_2^2}{2} ercf(\alpha_2) \right)\label{eqIIIE2}\\
\frac{\partial Q}{\partial t} &=& -\left( Q - \frac{2\epsilon E_2}{K_1K_2} \right) f_{R_1}(t)  -\left( Q - \frac{2\epsilon E_1}{K_1K_2} \right) f_{R_2}(t) \label{eqIIIQ}
\eeq
with the following definitions:

\beq 
R_1 &=& E_1 -\epsilon Q + \frac{\epsilon^2
E_2}{K_1K_2}~~~~~,~~~~~R_2 = E_2 -\epsilon Q + \frac{\epsilon^2
E_1}{K_1K_2},\label{eqIIIR1}\\ f_{R_i}(t) &=& \frac{K_i\Delta_i^2}{2}\beta
\exp \left(-\beta \frac{K_i\Delta_i^2}{2} (1-2R_i(t)\beta) \right)
erfc(\alpha_i (t) (4R_i(t)\beta - 1))~~~~{\rm with}~~~~~\alpha_i =
\sqrt{\frac{K_i\Delta_i^2}{16R_i}}~~~~,\label{eqIIIf} \eeq
where the error function was defined in (\ref{eqIIIerfc}).

Definitions (\ref{eqIIIf}) hold for $ i=1,2 $, each $i$
representing one of the two systems. Note that the whole dynamics 
is contained in the function $ f_{R_i(t)} $ .
In what follows we will be especially interested in the zero-temperature
case where relaxation time diverges and dynamics is slow and glassy. For
$T=0$ the function $f_{R_i(t)} $ in (\ref{eqIIIf}) becomes

\be
f_{R_i} = \frac{2\alpha_i}{\sqrt{\pi}} \exp(-\alpha_i^2)~~~.
\label{eqIIIfT0}
\ee

\subsection{Asymptotic long-time expansion for the one-time quantities}

The asymptotic solution of equations
(\ref{eqIIIE1},\ref{eqIIIE2},\ref{eqIIIQ}) may be guessed from the
behavior of the energy (\ref{APAEN}) for the uncoupled systems. Trying a
solution of the type

\be
E_1=\frac{a}{\log(t)}~~~~~,~~~~~E_2=\frac{b}{\log(t)}~~~~~,~~~~~Q=\frac{c}{\log(t)}~~~~~,
\label{eqIIIas}
\ee we can only solve the asymptotic behavior in the limit $ \e \approx
0. $ This is a consequence of the fact that the quantities $ R_1 $ and
$ R_2 $ are different in general and we have a system of four
equations with three parameters. There is not any general solution for
this system, but in the limit $ \e \approx 0 $ the quantities $ R_1 $
and $ R_2 $ become identical to the corresponding energies to leading order yielding only three
equations with three unknown parameters ($a,b,c$). In this limit the
value of the coefficients $a,b,c$ may be easily obtained yielding

\be
a=\frac{K_1^2K_2\Delta_1^2}{16(K_1K_2-\eps^2)}~~~~~,~~~~~b=\frac{K_1K_2^2\Delta_2^2}{16(K_1K_2-\eps^2)}~~~~~,~~~~~c=\frac{\eps
J}{8(K_1K_2-\eps^2)},\label{eqIIIa}\ee
where $J$ was defined in eq.(\ref{eqI}).At first
order in logarithmic corrections $1/\log(t)$ we find in the limit $ \e
\approx 0: $

\be
E_1=\frac{K_1^2K_2\Delta_1^2}{16(K_1K_2-\eps^2)\log(t)}~~~~~,~~~~~E_2=\frac{K_1K_2^2\Delta_2^2}{16(K_1K_2-\eps^2)\log(t)}~~~~~,~~~~~Q=\frac{\eps
J}{8(K_1K_2-\eps^2)\log(t)}\label{eqIIIaa}~~~~.  \ee

Note that, in the long time limit both energies $E_1,E_2$ tend to zero
logarithmically but their relative difference $\frac{E_1-E_2}{E_1}$
stays finite. In the limit of small coupling constant we can do a more
refined expansion yielding: \beq
E_1=\frac{K_1^2K_2\Delta_1^2}{16(K_1K_2-\eps^2)}\frac{1}{\log(t)+\frac{1}{2}\log(\log(t))}\label{eqIIIaaa}~~+~~
{\mathcal{O}}(\frac{1}{\log^2(t)}),\\
E_2=\frac{K_1K_2^2\Delta_2^2}{16(K_1K_2-\eps^2)}\frac{1}{\log(t)+\frac{1}{2}\log(\log(t))}\label{eqIIIbbb}~~+~~
{\mathcal{O}}(\frac{1}{\log^2(t)}),\\ Q=\frac{\eps
J}{8(K_1K_2-\eps^2)}\frac{1}{\log(t)+\frac{1}{2}\log(\log(t))}\label{eqIIIccc}~~+~~
{\mathcal{O}}(\frac{1}{\log^2(t)}), \eeq where we have put explicitly
the terms of order $ 1/\log^2(t) $ as sub-dominant corrections. These
terms come from the fact that the true expressions for the energies
and the overlap should be, in order to match the coefficients in
(\ref{eqIIIE1},\ref{eqIIIE2},\ref{eqIIIQ}):

\be
E_1=\frac{a}{\log(At)}~~~~~,~~~~~E_2=\frac{b}{\log(Bt)}~~~~~,~~~~~Q=\frac{c}{\log(Ct)},
\label{eqIIIaas}
\ee
which gives a correction of order  $ 1/\log^2(t) $ in expressions (\ref{eqIIIaa}).

\begin{figure}[tbp]
\begin{center}
\includegraphics*[width=9cm,height=5.5cm]{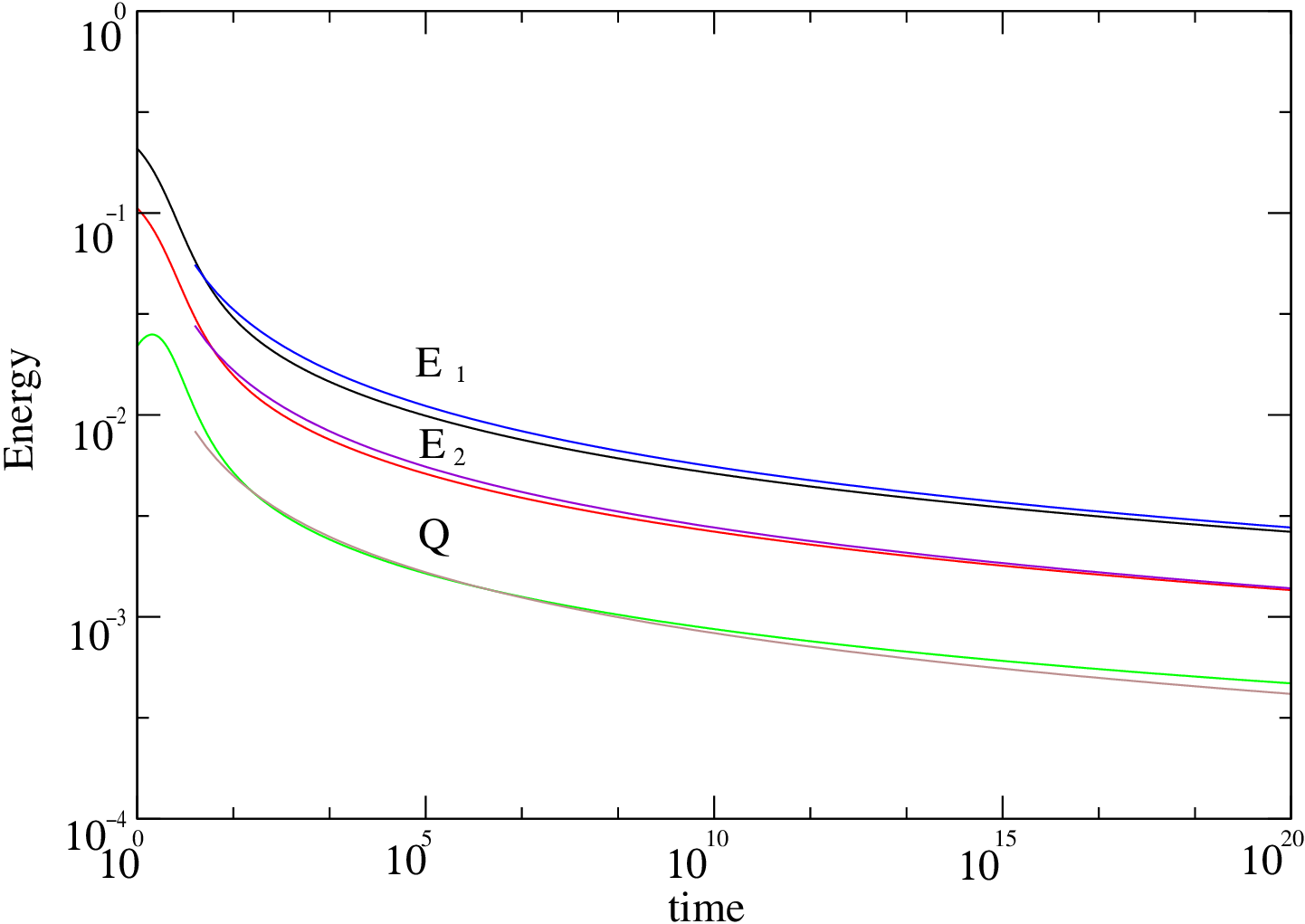}
\vskip 0.1in
\caption{The decay of the energies and the overlap for two systems with
$ K_1=2, K_2=1, \Delta_1=1, \Delta_2=1 $ and $ \eps=0.2 $. The longest
lines are the numerical solution for the dynamic equations, while the
shorter ones are the corresponding asymptotic behaviors.
\label{fig1}}
\end{center}
\end{figure}

In Fig.\ref{fig1} we show the evolution for the energies and the
overlap for two systems of harmonic oscillators with a small value of
$ \eps. $ We also show the asymptotic behavior
(\ref{eqIIIaaa},\ref{eqIIIbbb},\ref{eqIIIccc}). We can see that the
two energies remain different even at long times. We will see that
this feature is very important for describing the non-equilibrium
state of the whole system. We can also see that the asymptotic
expansions are in good agreement with the numerical solution of the
dynamic equations. Nevertheless, there are systematic deviations at long
times being consequence of the limited range of validity ($\eps\ll 1$)
of the asymptotic solution (\ref{eqIIIaaa},\ref{eqIIIbbb},\ref{eqIIIccc}). If $ K_1\Delta_1^2 =K_2\Delta_2^2 $ the energies of the two oscillators become identical (note that equations (\ref{energy},\ref{eqIIIerfc},\ref{ft}) only depend on the constant $  K\Delta^2 $) and asymptotic dynamics also.

\subsection{Correlations and responses.}

The set of equations for the four correlation and overlap functions
defined in (\ref{a},\ref{b}) can be written as:

\begin{equation}
\frac{\partial}{\partial t} \left( \begin{array}{c}
C_1(t,s) \\ C_2(t,s) \\ Q_1(t,s) \\ Q_2(t,s)
\end{array}\right) =
 -\left( \begin{array}{cccc}
f_{R_1}(t) & 0 & -\frac{\epsilon}{K_1}f_{R_1}(t) & 0 \\
0 & f_{R_2}(t) & 0 &  -\frac{\epsilon}{K_2}f_{R_2}(t) \\
 -\frac{\epsilon}{K_2}f_{R_2}(t) & 0 &  f_{R_2}(t) & 0 \\
0 &  -\frac{\epsilon}{K_1}f_{R_1}(t) & 0 & f_{R_1}(t) \end{array}\right)
\left( \begin{array}{c}
C_1(t,s) \\ C_2(t,s) \\ Q_1(t,s) \\ Q_2(t,s) \end{array} \right) 
\label{eqIIImatrix}
\end{equation}

with the subsidiary boundary conditions

\begin{equation}
C_1(s,s)=\frac{2E_1(s)}{K_1}~~~~~,~~~~~C_2(s,s)=\frac{2E_2(s)}{K_2}~~~~~,~~~~~Q_1(s,s)=Q_2(s,s)=Q(s)~~~~~.
\label{ci1}
\end{equation}

The equilibrium solution can be appropriately worked out because the
matrix coefficients are time-independent. If we write the matrix equation
in compact form

\be
\frac{\partial \vec{C}}{\partial t}=M \vec{C}
\label{eqIIIM}\ee

\noindent 
with $\vec{C}=(C_1,C_2,Q_1,Q_2)$ the solution is

\be
\vec{C}(t,s)=\vec{C}(s,s)\exp\bigl(M_{\rm eq}(t-s)\bigr)~~~~~.
\label{eqIIIC}
\ee

The precise results for correlations and overlaps are reported in the
Appendix B (formulae (\ref{ApeqC1}-\ref{finale})), the initial
conditions being given in (\ref{ci1}). In the
non-equilibrium case it is not possible to write down an exact solution
for equation (\ref{eqIIIM}) for any value of $\epsilon$. The formal
solution of (\ref{eqIIIM}) is

\be
\vec{C}(t,s)={\cal T}\exp\bigl(\eps\int_{s}^tB_I(s')\bigr)\vec{C}(s,s)~~~~~,
\label{eqIIITC}
\ee
which can be worked out perturbatively up to any order around
$\eps=0$ ($\cal T $ stands for the time ordered product). In the Appendix C, we give some details how to construct such
expansion. Up to order $\eps^2$ the solution for the components
$C_1,C_2$ of the four component vector $\vec{C}$ are

\beq C_1(t,s)= \exp \Bigl( - \int_s^t f_{R_1}(x)dx \Bigr) \Bigl(
\frac{2E_1(s)}{K_1} + \e \frac{Q(s)}{K_1} \int_s^t dt'f_{R_1}(t') e^{
\int_s^{t'}(f_{R_1}(x)-f_{R_2}(x))dx}+\nonumber\\ 
\e^2 \frac{2E_1(s)}{K_2 K_1^2} \int_s^t dt' \int_s^{t'} dt''
f_{R_1}(t')f_{R_2}(t'') e^{ \int_{t''}^{t'}(f_{R_1}(x)-f_{R_2}(x))dx}
\Bigr),\label{eqIIIC1eps} \eeq

\beq C_2(t,s)= \exp \Bigl( - \int_s^t f_{R_2}(x)dx \Bigr) \Bigl(
\frac{2E_2(s)}{K_2} + \e \frac{Q(s)}{K_2} \int_s^t dt'f_{R_2}(t') e^{
\int_s^{t'}(f_{R_2}(x)-f_{R_1}(x))dx} +\nonumber\\ 
\e^2 \frac{2E_2(s)}{K_1 K_2^2} \int_s^t dt'
\int_s^{t'} dt'' f_{R_2}(t')f_{R_1}(t'') e^{
\int_{t''}^{t'}(f_{R_2}(x)-f_{R_1}(x))dx} \Bigr). \label{eqIIIC2eps}\eeq

Similar expansions are obtained from $Q_1(t,s),Q_2(t,s)$. Here we do not
report them because correlations are enough to
analyze the effective temperatures.  Similarly we can also obtained
expressions for the responses as detailed in the Appendix B. The time
evolution for the four possible response functions is given by 

\beq
\frac{\partial G_1(t,s)}{\partial t}&=&- \left( G_1(t,s)f_{R_1}(t) -
\frac{f_{R_1}(t)}{K_1} \delta(t-s) -\frac{\epsilon}{K_1} G_2'(t,s)
f_{R_1}(t) \right) \label{eqIIIG1} \\ \frac{\partial G_2(t,s)}{\partial
t}&=&- \left( G_2(t,s)f_{R_2}(t) - \frac{f_{R_2}(t)}{K_2} \delta(t-s)
-\frac{\epsilon}{K_2} G_1'(t,s) f_{R_2}(t) \right) \label{eqIIIG2} \\
\frac{\partial G_1'(t,s)}{\partial t}&=&- \left( G_1'(t,s)f_{R_1}(t)
-\frac{\epsilon}{K_1} G_2(t,s) f_{R_1}(t) \right)\label{eqIIIG1p} \\
\frac{\partial G_2'(t,s)}{\partial t}&=&- \left(
G_2'(t,s)f_{R_2}(t)-\frac{\epsilon}{K_2} G_1(t,s) f_{R_2}(t) \right)~~~~~.
\label{eqIIIG2p}\eeq

As explained in Appendix B these equations must be solved with
the subsidiary boundary conditions,

\beq G_1(s,s) = \frac{ f_{R_1}(s)}{K_1}~~~~~;~~~~~G_2(s,s) = \frac{
f_{R_2}(s)}{K_2}~~~~~,~~~~~G_1'(s,s) = 0~~~~~,~~~~~G_2'(s,s) =
0\label{eqIIIiniG}~~~.  \eeq

The initial conditions for $G_1,G_2$ come from the delta-terms in their
equations. The other two initial conditions for $G_1',G_2'$ come from
the fact that there is no discontinuous jump in the response function of
one system when we apply the field to the other system. This result also
holds in the framework of the Langevin dynamics and manifests in the
equations for the magnetizations (see in Appendix B
(\ref{eqApM1},\ref{eqApM2})) as the absence of a field $h_2$ in the
equation for $M_1$ and the absence of a term $h_1$ in the equation for
$M_2$.

In equilibrium the expressions for the responses $G_1,G_2$ are given in the
Appendix B. Up to order $\eps^2$ the expression for the off-equilibrium responses
$G_1,G_2$ can be solved analogously as done for the correlations and are given by

\be G_1(t,s)= \frac{f_{R_1}}{K_1} \exp \bigl( \int_s^t f_{R_1}(x)dx
\bigr) \Bigl (1 +  \frac{\e^2}{K_1K_2} \int_s^t dt' \int_s^{t'} dt''
f_{R_1}(t')f_{R_2}(t'') e^{ \int_{t''}^{t'}(f_{R_1}(x)-f_{R_2}(x))dx}
\Bigr) \label{eqIIIG1eps}\ee

\be G_2(t,s)= \frac{f_{R_2}}{K_2} \exp \bigl( \int_s^t f_{R_2}(x)dx
\bigr) \Bigl(1+ \frac{\e^2}{K_1K_2} \int_s^t dt' \int_s^{t'}
dt'' f_{R_2}(t')f_{R_1}(t'') e^{
\int_{t''}^{t'}(f_{R_2}(x)-f_{R_1}(x))dx} \Bigr)~~~~~.
\label{eqIIIG2eps}\ee

\section{The dynamically coupled (or parallel) case}

For the dynamically coupled case the calculations proceed similarly as
to the previous dynamically uncoupled case.  The evolution equations for the overlap and the energies
(\ref{q}) are: \beq \frac{\partial Q}{\partial t}&=& - \left( Q
-\frac{\epsilon I}{J} \right) f_{E_N}(t)
\label{c2q}\\
\frac{\partial E_1}{\partial t} &=& \frac{K_1\Delta_1^2}{4} erfc \left( \frac{J}{\sqrt{4E_N}} \right) +\left( \frac{K_1\Delta_1^2}{2\beta} - 2E_1\Delta_1^2 K_1 + \epsilon Q\Delta_1^2 K_1 \right) \frac{f_{E_N}(t)}{2J}
\label{c2e1}\\
\frac{\partial E_2}{\partial t} &=& \frac{K_2\Delta_2^2}{4} erfc \left( \frac{J}{\sqrt{4E_N}} \right) +\left( \frac{K_2\Delta_2^2}{2\beta} - 2E_2\Delta_2^2 K_2 + \epsilon Q\Delta_2^2 K_2 \right) \frac{f_{E_N}(t)}{2J}~~~~~.
\label{c2e2}
\end{eqnarray}       
$I$ and $J$ were defined in (\ref{eqI}) and the new
quantities $E_N$, $f_{E_N}$ and $E_T$ ($E_T$ is not the total energy) are given by

\beq
E_N &=& E_T - 2\epsilon QJ +\epsilon^2 I~~~~~,~~~~~E_T= E_1K_1\Delta_1^2 + E_2K_2\Delta_2^2\label{eqIVET}\\
f_{E_N}(t)&=&J\beta \exp \left( \beta^2 E_N-J \beta \right) erfc\left( \frac{2E_N\beta - J}{\sqrt{4E
_N}} \right)\label{eqIVf}
\eeq

\noindent and the expression (\ref{eqIVf}) at zero-temperature is

\be
f_{E_N}(t)=\frac{J}{\sqrt{E_N\pi}}\exp(-\frac{J^2}{4E_N})~~~~.
\ee

\subsection{Asymptotic long-time expansion for the one-time quantities}

Proceeding similarly as done in the former section we can find the
asymptotic expressions for the energies and overlaps. In this case we
can find a solution for $ \eps $ finite due to the fact that in this
case we have only one dynamic function $ f_{E_N} .$ We find, to leading order in $1/\log(t)$ 

\be
E_1=\frac{K_1K_2J}{8(K_1K_2-\eps^2)\log(t)}~~~~~,~~~~~E_2=\frac{K_1K_2J}{8(K_1K_2-\eps^2)\log(t)}~~~~~,~~~~~Q=\frac{\eps J}{4(K_1K_2-\eps^2)\log(t)}\label{eqIVaa}~~~~.  \ee

Note that, contrarily to results
(\ref{eqIIIaa}) for the dynamically
uncoupled case, the energies $E_1,E_2$ asymptotically coincide and the
relative difference $(E_1-E_2)/E_1$ vanishes in the long-time
limit. This difference of behaviors is not casual and has a physical
interpretation that we will discuss later. The more precise expansion
turns out to be,

\beq
E_1=\frac{K_1K_2J}{8(K_1K_2-\eps^2)}\frac{1}{\log(t)+\frac{1}{2}\log(\log(t))}~~+~~
{\mathcal{O}}(\frac{1}{\log^2(t)})\label{eqIVaaa}\\E_2=\frac{K_1K_2J}{8(K_1K_2-\eps^2)}\frac{1}{\log(t)+\frac{1}{2}\log(\log(t))}~~+~~
{\mathcal{O}}(\frac{1}{\log^2(t)})\label{eqIVbbb}\\Q=\frac{\eps
J}{4(K_1K_2-\eps^2)}\frac{1}{\log(t)+\frac{1}{2}\log(\log(t))}~~+~~
{\mathcal{O}}(\frac{1}{\log^2(t)})\label{eqIVccc}~~~~~. \eeq

\begin{figure}[tbp]
\begin{center}
\includegraphics*[width=9cm,height=5.5cm]{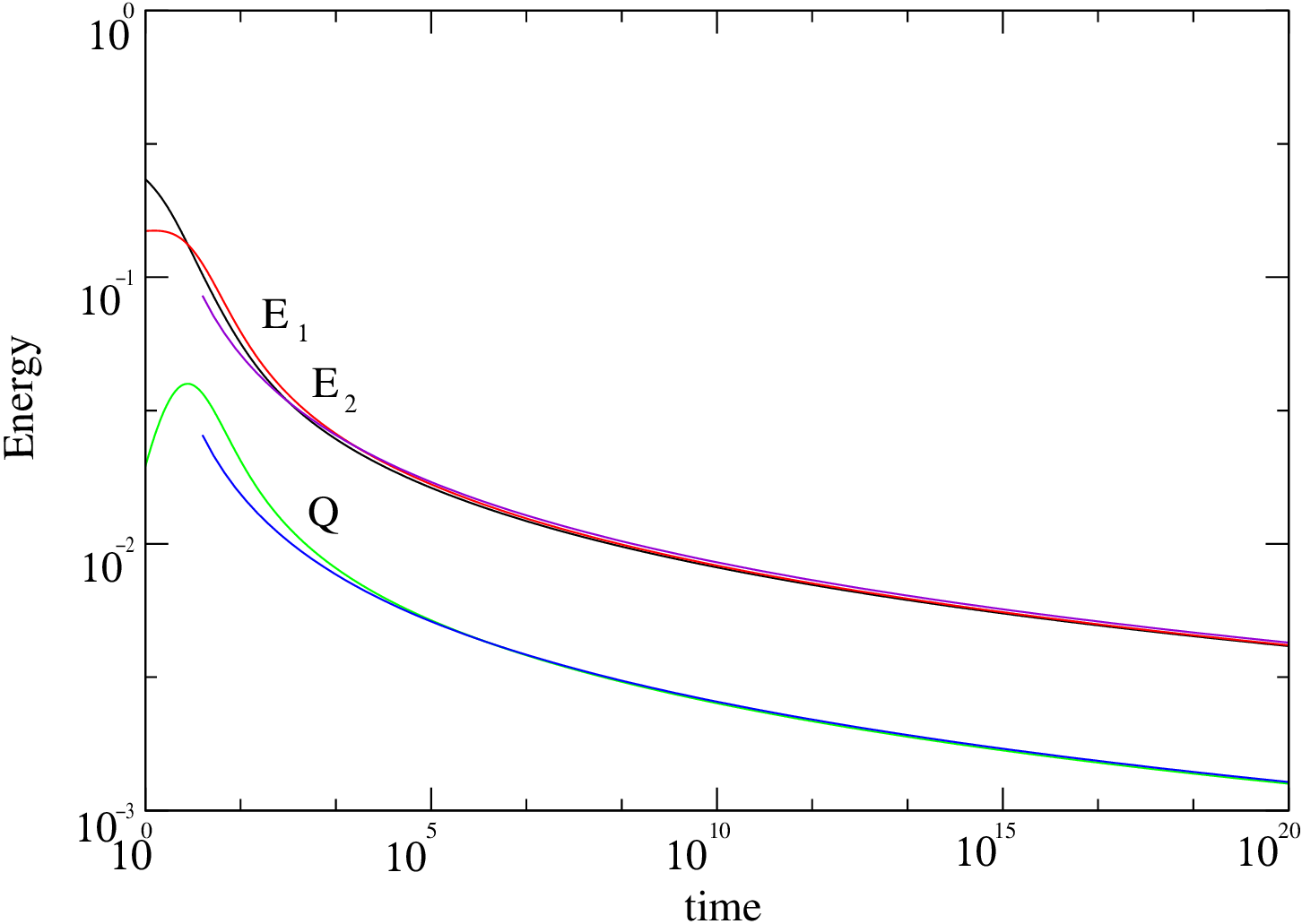}
\vskip 0.1in
\caption{The decay of the energies and the overlap for two systems
with $ K_1=2, K_2=1, \Delta_1=1, \Delta_2=1 $ and $ \eps=0.3 $. Black
lines are the numerical solution for the dynamic equations, while the
blue ones are the corresponding asymptotic behaviors.
\label{fig2}}
\end{center}
\end{figure}

Let us stress that, contrarily to the dynamically uncoupled or
sequential case the previous expressions are valid to any order in $\eps$.
The origin of the $ 1/\log^2(t) $ terms in previous expressions is the
same as in the uncoupled case
(\ref{eqIIIaaa},\ref{eqIIIbbb},\ref{eqIIIccc}). In Fig.~\ref{fig2}, we
show the numerical solution for the evolution of the energies and the
overlap as well as the asymptotic expansions
(\ref{eqIVaaa},\ref{eqIVbbb},\ref{eqIVccc}).

We have said that the relative difference $(E_1-E_2)/E_1$ vanishes in
the long-time limit. It is not difficult to see how this happens. The
time-evolution for the quantity $E_1/E_2$ is easy to derive from
eqs.(\ref{c2q},\ref{c2e1},\ref{c2e2}) in the asymptotic long-time
limit $E_N\to 0$. One then finds the following expansion to leading
order

\be
\frac{E_1}{E_2}=1-\frac{K_1\Delta_1^2-K_2\Delta_2^2}{2J\log(t)}~~~.
\label{E1divE2}
\ee
If $K_1\Delta_1^2 =K_2\Delta_2^2$ the correction is even smaller. 

Interestingly, this leading correction does not depend on $\epsilon$
showing that the two energies $E_1,E_2$ approach each other at a rate
determined by the fact that the whole dynamics of the model is coupled
and not by the fact that the two oscillator systems are coupled by the
presence of a term $\eps Q$ in the Hamiltonian.  The behavior of
this quantity is shown in figure~(\ref{fig3}) for different values of
$\epsilon$ together with the asymptotic
expansion (\ref{E1divE2}).

\begin{figure}[tbp]
\begin{center}
\includegraphics*[width=9cm,height=5.5cm]{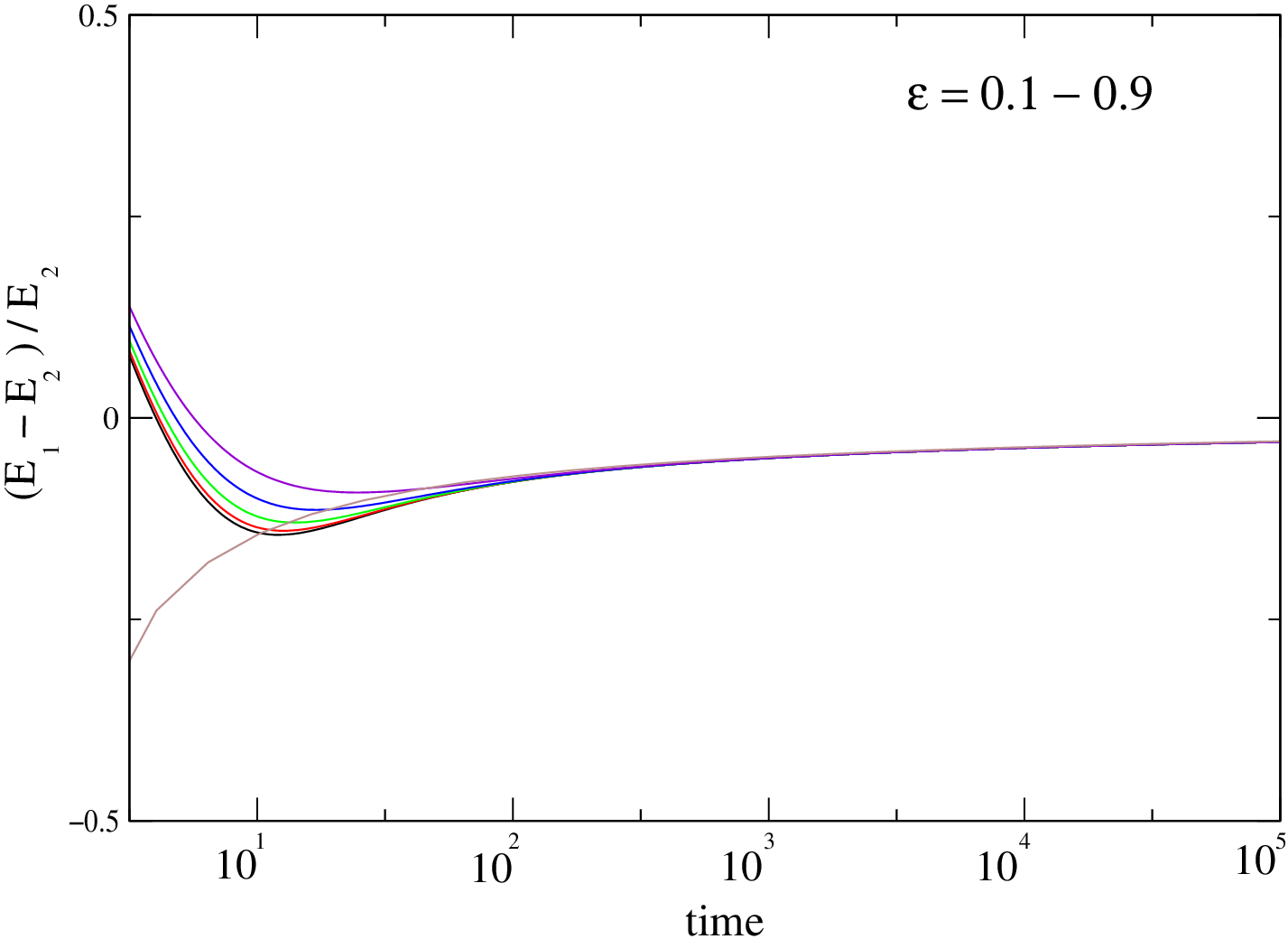}
\vskip 0.1in
\caption{Relative energy difference $E_1/E_2-1$ for two systems with $
K_1=2, K_2=1, \Delta_1=1, \Delta_2=1 $ and $ \eps=0.1,0.3,0.5,0.7,0.9 $
(from top to bottom). The asymptotic
prediction (\ref{E1divE2}) is also shown.\label{fig4}}
\end{center}
\end{figure}

\subsection{Correlations and responses.}

Following similar methods as for the dynamically uncoupled case
presented before we can write down the equations for correlations
and overlaps

\begin{equation}                 
\frac{\partial C_1(t,s)}{\partial t} = - \left( K_1\Delta_1^2 C_1(t,s) - \epsilon \Delta_1^2  Q_1(t,s) \right) \frac{f_{E_N}(t)}{2J} \label{eqIVC1}
\end{equation}  
\begin{equation}                 
\frac{\partial C_2(t,s)}{\partial t} = - \left( K_2\Delta_2^2 C_2(t,s) - \epsilon \Delta_2^2  Q_2(t,s) \right) \frac{f_{E_N}(t)}{2J} \label{eqIVC2}
\end{equation} 
\begin{equation}                 
\frac{\partial Q_1(t,s)}{\partial t} = - \left( K_2\Delta_2^2 Q_1(t,s) - \epsilon \Delta_2^2  C_1(t,s) \right) \frac{f_{E_N}(t)}{2J} \label{eqIVQ1}
\end{equation} 
\begin{equation}                 
\frac{\partial Q_2(t,s)}{\partial t} = - \left( K_1\Delta_1^2 Q_2(t,s) - \epsilon \Delta_1^2  C_2(t,s) \right) \frac{f_{E_N}(t)}{2J}~~~~~, \label{eqIVQ2}
\end{equation} 
with the subsidiary boundary conditions given by equations
(\ref{ci1}).

In matrix form these equations reduce
to the equation (\ref{eqIIIM}).  As explained for the uncoupled case,
this set of equations can be exactly solved only in the equilibrium
regime where the coefficients are time-independent. The solution is then
given by the equation (\ref{eqIIIC}), the expressions for correlations
and overlaps are the formulae (\ref{ApeqC1},\ref{ApeqC2}) with
$a=-\frac{K_1\Delta_1^2}{2J} f_{E_N};b=\frac{\epsilon\Delta_1^2}{2J}
f_{E_N}; c=\frac{\epsilon\Delta_2^2}{2J} f_{E_N};
d=-\frac{K_2\Delta_2^2}{2J} f_{E_N}$
and the formulae (\ref{ApeqQ1},\ref{ApeqQ2}) with
$a=-\frac{K_2\Delta_2^2}{2J} f_{E_N}; b=\frac{\epsilon\Delta_2^2}{2J}
f_{E_N}; c=\frac{\epsilon\Delta_1^2}{2J}  f_{E_N};
d=-\frac{K_1\Delta_1^2}{2J} f_{E_N}$. 

In the most general case where the coefficients of the matrix equation
are time dependent the exact solution can be written in the closed form
(\ref{eqIIITC}) which can be expanded to any order in $\e$ as explained
in the Appendix C. As in the uncoupled case we present here the results
up to order $\eps^2$ only for the correlations,

\beq C_1(t,s)= \exp \left( -\frac{K_1\Delta_1^2}{2J} \int_s^t
f_{E_N}(x)dx \right) \Bigl( \frac{2E_1(s)}{K_1} + \e
\frac{\Delta_1^2Q(s)}{2J} \int_s^t dt'f_{E_N}(t') e^{\left(
\frac{K_1\Delta_1^2-K_2\Delta_2^2}{2J} \right) \int_s^{t'}f_{E_N}(x)dx}
+ \nonumber \\ \e^2 \frac{2E_1(s)\Delta_1^2\Delta_2^2}{4J^2 K_1} \int_s^t
dt' \int_s^{t'} dt'' f_{E_N}(t')f_{E_N}(t'') e^{\left(
\frac{K_1\Delta_1^2-K_2\Delta_2^2}{2J} \right)
\int_{t''}^{t'}f_{E_N}(x)dx} \Bigr) \label{eqIVC1eps}\\
C_2(t,s)= \exp \left( -\frac{K_2\Delta_2^2}{2J} \int_s^t f_{E_N}(x)dx
\right) \Bigl( \frac{2E_2(s)}{K_2} + \e \frac{\Delta_2^2Q(s)}{2J} \int_s^t
dt'f_{E_N}(t') e^{\left( \frac{K_2\Delta_2^2-K_1\Delta_1^2}{2J} \right)
\int_s^{t'}f_{E_N}(x)dx} \nonumber +\\ \e^2
\frac{2E_2(s)\Delta_2^2\Delta_1^2}{4J^2 K_2} \int_s^t dt' \int_s^{t'}
dt'' f_{E_N}(t')f_{E_N}(t'') e^{\left(
\frac{K_2\Delta_2^2-K_1\Delta_1^2}{2J} \right)
\int_{t''}^{t'}f_{E_N}(x)dx}\Bigr)~~~~~.\label{eqIVC2eps}
\end{eqnarray}

Responses $G_{1,2},G'_{1,2}$ can be worked out in a similar way as shown
in the Appendix B for the uncoupled case,

\beq \frac{\partial G_1(t,s)}{\partial t}=-\frac{K_1\Delta_1^2}{2J}
\left( G_1(t,s)f_{E_N}(t) - \frac{f_{E_N}(t)}{K_1} \delta(t-s)
-\frac{\epsilon}{K_1} G_2'(t,s) f_{E_N}(t) \right)\label{eqIIG1}  \\
\frac{\partial G_2(t,s)}{\partial t}=-\frac{K_2\Delta_2^2}{2J} \left(
G_2(t,s)f_{E_N}(t) - \frac{f_{E_N}(t)}{K_2} \delta(t-s)
-\frac{\epsilon}{K_2} G_1'(t,s) f_{E_N}(t) \right)\label{eqIIG2}\\
\frac{\partial G_1'(t,s)}{\partial t}=-\frac{K_1\Delta_1^2}{2J} \left( G_1'(t,s)f_{E_N}(t) -\frac{\epsilon}{K_1} G_2(t,s) f_{E_N}(t) \right) \label{eqIIG1p}\\ 
\frac{\partial G_2'(t,s)}{\partial t}=-\frac{K_2\Delta_2^2}{2J} \left( G_2'(t,s)f_{E_N}(t)-\frac{\epsilon}{K_2} G_1(t,s) f_{E_N}(t) \right)~~~~~,\label{eqIIG2p}
\eeq

with the subsidiary boundary conditions (note that for $G_1,G_2$ they
are different from those in (\ref{eqIIIiniG})),

\be
G_1(s,s) = \frac{\Delta_1^2 f_{E_N}(s)}{2J}~~~~,~~~~G_2(s,s) = \frac{\Delta_2^2 f_2(s)}{2J}~~~~,~~~~G_1'(s,s) = 0~~~~,~~~~G_2'(s,s) = 0~~~~.
\ee

It is a simple exercise to check in equilibrium whether these responses give
the correct value of the susceptibility
(\ref{ji}). In equilibrium responses only depend on
the difference of times. As the susceptibility is just the integral of the
response function we can integrate the equations (for simplicity we
shall consider that $s=0$). Then, the equilibrium susceptibility of
every system is just:

\be
\chi = \int_0^\infty G(t) dt~~~~~.\label{chi}
\ee
Integrating the equations for response functions we obtain:
\beq
G_1(\infty) - G_1(0) &=& -\frac{f_{E_N}K_1\Delta_1^2}{2J} \left( \chi_1 + \frac{\epsilon}{K_1} \chi_2' \right) = -\frac{f_{E_N}K_1\Delta_1^2}{2JK_1}\\ 
G_2'(\infty) - G_2'(0) &=&  -\frac{f_{E_N}K_2\Delta_2^2}{2J} \left( \chi_2' + \frac{\epsilon}{K_2} \chi_1 \right) = 0\\
G_2(\infty) - G_2(0) &=& -\frac{f_{E_N}K_2\Delta_2^2}{2J} \left( \chi_2 + \frac{\epsilon}{K_2} \chi_1' \right) = -\frac{f_{E_N}K_2\Delta_2^2}{2JK_2}\\ 
G_1'(\infty) - G_1'(0) &=&  -\frac{f_{E_N}K_1\Delta_1^2}{2J} \left( \chi_1' + \frac{\epsilon}{K_1} \chi_2 \right) = 0~~~~~.
\eeq

At very long times, ergodicity imposes
$G_{1,2}(\infty)=G_{1,2}'(\infty)=0$. These equations give the exact
results for the equilibrium susceptibilities eqs.(\ref{ji}).
In equilibrium the responses can be easily computed and one gets (to
keep formulae at minimum we only report the results for $G_1$ and $G_2$),

\beq
G_1(t,s) = \frac{\Delta_1^2 f_{E_N}(s)}{2J}\left( \frac{\lambda_2 - a}{\lambda_2-\lambda_1} \exp(\lambda_1 (t-s)) -\frac{\lambda_1 - a}{\lambda_2-\lambda_1} \exp(\lambda_2 (t-s)) \right)\label{eqIVG1eq}\\ 
G_2(t,s) =   \frac{\Delta_2^2 f_{E_N}(s)}{2J}\left( \frac{\lambda_2 - a}{\lambda_2-\lambda_1} \exp(\lambda_1 (t-s)) -\frac{\lambda_1 - a}{\lambda_2-\lambda_1} \exp(\lambda_2 (t-s)) \right)~~~~~, \label{eqIVG2eq}
\eeq

with the usual expression (\ref{eigen2}) for
 $\lambda_1,\lambda_2$. For $G_1$
 the values of the constants are:

\begin{equation}
a=-\frac{K_1\Delta_1^2}{2J} f_{E_N}; ~~~~~
b=\frac{\epsilon\Delta_1^2}{2J}  f_{E_N}; ~~~~~
c=\frac{\epsilon\Delta_2^2}{2J}  f_{E_N}; ~~~~~
d=-\frac{K_2\Delta_2^2}{2J} f_{E_N}\label{eqIVc}
\end{equation}
while for $G_2$ the same results (\ref{eqIVc}) are valid but
interchanging the indices 1 and 2.

In the general off-equilibrium case the result for $G_1,G_2,G'_1,G'_2$
can be worked out perturbatively. Here we only write the expression up
to order $\eps^2$

\beq G_1(t,s)= \frac{f_{E_N}\Delta_1^2}{2J} \exp \Bigl(
-\frac{K_1\Delta_1^2}{2J} \int_s^t f_{E_N}(x)dx \Bigr)\times \nonumber \\ \Bigl( 1+\e^2\frac{\Delta_1^2\Delta_2^2}{4J^2} \int_s^t dt' \int_s^{t'} dt''
f_{E_N}(t')f_{E_N}(t'') e^{\bigl( \frac{K_1\Delta_1^2-K_2\Delta_2^2}{2J}
\bigr) \int_{t''}^{t'}f_{E_N}(x)dx} \Bigr)\label{eqIVG1eps} \\
G_2(t,s)= \frac{f_{E_N}\Delta_2^2}{2J} \exp \Bigl(
-\frac{K_2\Delta_2^2}{2J} \int_s^t f_{E_N}(x)dx \Bigr)\times \nonumber \\
 \Bigl( 1+ \e^2 \frac{\Delta_2^2\Delta_1^2}{4J^2} \int_s^t dt' \int_s^{t'} dt'' f_{E_N}(t')f_{E_N}(t'') e^{\bigl( \frac{K_2\Delta_2^2-K_1\Delta_1^2}{2J} \bigr) \int_{t''}^{t'}f_{E_N}(x)dx} \Bigr)\label{eqIVG2eps}
\eeq

\section{Results and discussion}

First of all we can check that the equilibrium results are the expected ones. It is easy to prove that, independent of the dynamics, the effective temperatures are just the temperature of the bath:
\be
T_{\rm eff}^1 = 2E_1 - \epsilon Q = T,~~~
T_{\rm eff}^2 = 2E_2 - \epsilon Q = T.
\ee
Because in equilibrium the energies of the subsystems are the same (see (\ref{eqQ})).

\subsection{Sequential case} 

In the off-equilibrium case the results are more interesting. It is easy
to verify the following expressions for (\ref{T1}) up to order $ \eps^2 $ :

\be
T_{\rm eff}^1=2E_1(s) + \frac{2}{f_{R_1}(s)} \frac{\partial E_1(s)}{\partial s} - \e Q(s) + \bigl( \eps Q(s) - \eps^2\frac{2E_1(s)}{K_1K_2} \bigr) \frac{f_{R_2}(s)}{f_{R_1}(s)}\int_s^t dt'f_{R_1}(t')e^{\int_s^{t'}(f_{R_1}(x)- f_{R_2}(x))dx} \label{eqIIITEFF1}
\end{equation}
\be
T_{\rm eff}^2=2E_2(s) + \frac{2}{f_{R_2}(s)} \frac{\partial E_2(s)}{\partial s} - \e Q(s) + \bigl( \eps Q(s) - \eps^2\frac{2E_2(s)}{K_1K_2} \bigr) \frac{f_{R_1}(s)}{f_{R_2}(s)}\int_s^t dt'f_{R_2}(t')e^{\int_s^{t'}(f_{R_2}(x)- f_{R_1}(x))dx}\label{eqIIITEFF2}~~~~~. 
\end{equation}

The expression for the total effective temperature is just
\be
T_{\rm eff}^T= \frac{\frac{\partial C_1(t,s)}{\partial s}+\frac{\partial C_1(t,s)}{\partial s}}{G_1 (t,s)+G_2(t,s)}~~~~~,\label{eqIIITEFF3} 
\end{equation}
where the correlations are given by expressions (\ref{eqIIIC1eps},\ref{eqIIIC2eps}) and the responses are given by (\ref{eqIIIG1eps},\ref{eqIIIG2eps}).   

From the equations (\ref{eqIIITEFF1},\ref{eqIIITEFF2}) we can see
immediately that the effective temperatures are well defined in the
regime in which the ratio $ \frac{t}{s} $ is finite. Otherwise, the last
term in the right hand side of (\ref{eqIIITEFF1}) and (\ref{eqIIITEFF2})
would diverge. At zero temperature and up to order $ \eps^2 $ in the
coupling constant, we have found that $E_1,E_2,Q$ decrease
logarithmically implying that both $ f_{R_1} $ and $ f_{R_2} $ decay
like $ 1/t $. Now let us consider $t,s$ both large but $t-s\ll s$. For a weak coupling (i.e $ \eps
\approx 0) $ the value of the effective temperatures are, in the limit $
s \rightarrow \infty $ but with $\frac{t}{s} $ finite:
   
\be
T_{\rm eff}^1\approx 2E_1(s) + {\mathcal{O}}(\eps^2) \approx \frac{K_1^2K_2\Delta_1^2}{8(K_1K_2-\eps^2)\log(t)} \label{D11}
\ee
\be
T_{\rm eff}^2\approx 2E_2(s) + {\mathcal{O}}(\eps^2) \approx \frac{K_1K_2^2\Delta_2^2}{8(K_1K_2-\eps^2)\log(t)}~~~~~.\label{D12}
\ee

This yields in the $s\to\infty$ limit a non vanishing relative
difference $T_{\rm eff}^1/T_{\rm eff}^2-1$. This is a consequence of the fact that the two energies are different in the long-time regime. Note that each effective temperature verifies the equipartition theorem in the limit of long
times as expected. The physical interpretation is clear: each system
is relaxing towards its equilibrium state slowly and at any time we
can consider that the systems are at ``quasi-equilibrium'' at their
corresponding effective temperatures. Obviously the concept of
``quasi-equilibrium'' is meaningful in a time window smaller than the
characteristic time-scale in which the system relaxes (i.e. during this
time-scale the effective temperatures do not change), hence we need to
impose that $t/s$ is finite.

\begin{figure}[tbp]
\begin{center}
\includegraphics*[width=9cm,height=5.5cm]{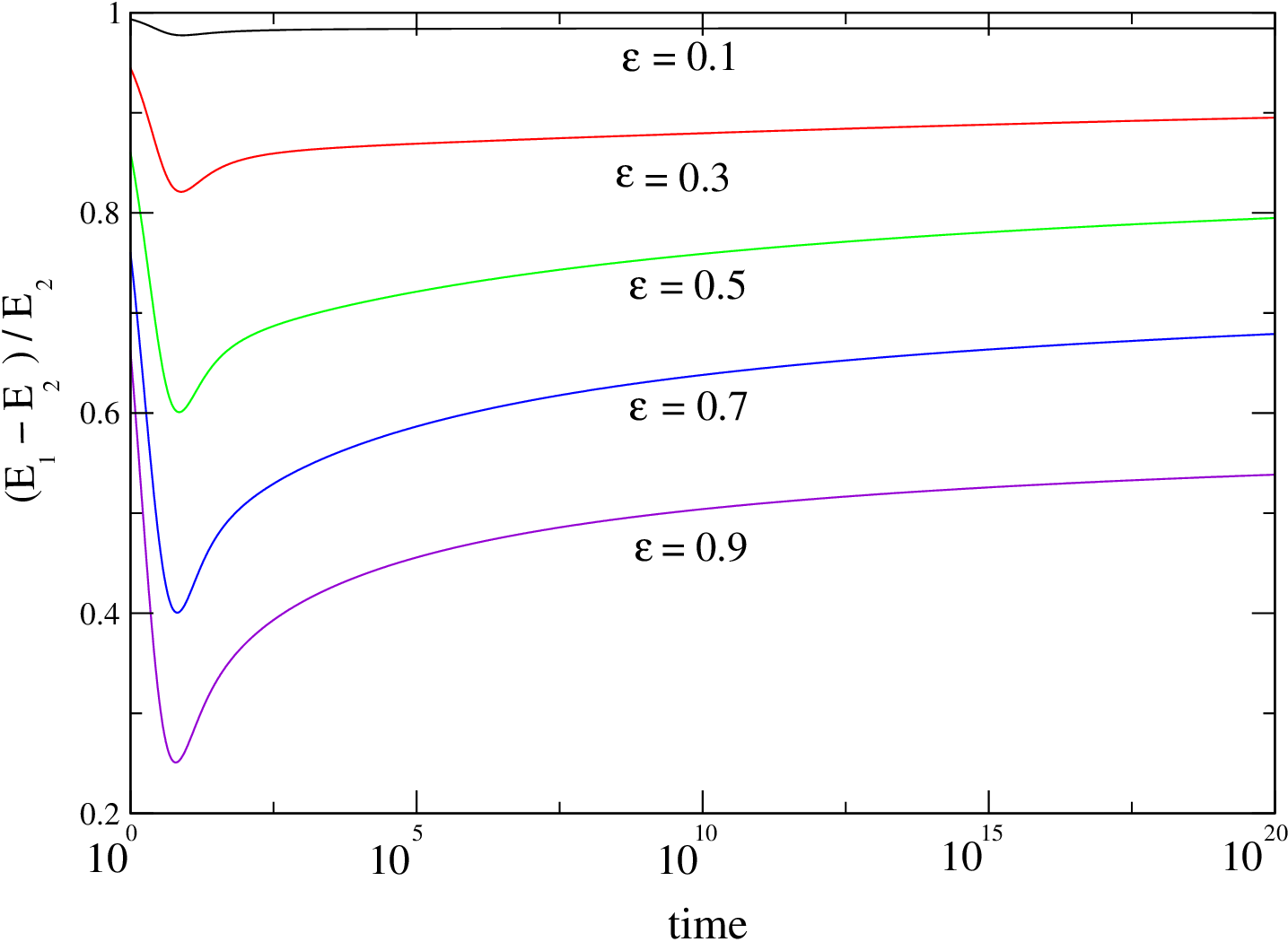}
\vskip 0.1in
\caption{Relative energy difference $E_1/E_2-1$ for two systems with $
K_1=2, K_2=1, \Delta_1=1, \Delta_2=1 $ and different values of $
\eps=0.1,0.3,0.5,0.7,0.9$. Note that the relative difference
increases with time.\label{fig3}}
\end{center}
\end{figure}

Let's think now about the global system. As we have seen, the energies
for the two systems remain different even at infinite times. This can
be explicitly seen in figure (\ref{fig3}) where we show how the
relative difference between the energies (or the effective
temperatures according to (\ref{D11},\ref{D12})) increases
monotonically as a function of time (for late times) for any value of
$\eps$. We may then conclude that a coupling in the Hamiltonian is not
enough to reach an equalization of effective temperatures.

This difference of the two effective temperatures implies that there
are some degrees of freedom hotter than others. One can then imagine
that there is always some kind of heat transfer or current flow going
from the ``hot degrees'' of freedom to the ``cold'' ones. Then, one
may ask why the effective temperatures do not asymptotically
equalize. The reason is that the off-equilibrium conductivity may
vanish with time fast enough for the heat transfer not to be able to
compensate such difference. In this situation, if we now compute the
total effective temperature (\ref{eqIIITEFF3}) for the whole system we
see that in the off-equilibrium regime this temperature does not
coincide with the sum of the energies of the systems. This fact
fortifies the definition of the effective temperature using the FDR
(\ref{eqX}) in off-equilibrium systems. For two systems in ``local''
equilibrium at two different temperatures, despite the fact that each
system verifies FDT, the sum of the two systems never verifies FDT
unless the two temperatures are equal. In our case, we have two
systems which are in ``quasi-equilibrium'' at two different effective
temperatures, so the $ T_{\rm eff}^T $ would never be the sum of the
two energies unless the two effective temperatures $T_{\rm
eff}^1,T_{\rm eff}^2$ were the same. In other words, two systems
thermodynamically stable at different temperatures are not globally
stable when put in contact.

\subsection{Parallel dynamics}

The effective temperatures (\ref{T1}) can be
exactly computed to order $\eps^2$. In the equilibrium regime, both the 
full expression derived from
(\ref{ApeqC1},\ref{ApeqC2},\ref{eqIVG1eq},\ref{eqIVG2eq}) and the
general approximate solutions 
(\ref{eqIVC1eps},\ref{eqIVC2eps},\ref{eqIVG1eps},\ref{eqIVG2eps}) up to
order $\eps^2$ computed in the equilibrium regime yield the bath
temperature for the three effective temperatures. In the non-equilibrium
case, up to order $\eps^2$, the results are:

\be
T_{\rm eff}^1=2E_1(s) + \frac{4J}{K_1\Delta_1^2f_{E_N}(s)} \frac{\partial E_1(s)}{\partial s} - \e Q(s) + \bigl( \eps Q(s) - \eps^2\frac{2E_1(s)}{K_1K_2} \bigr) K_2\Delta_2^2\int_s^t dt'f_{E_N}(t')e^{\frac{K_1\Delta_1^2-K_2\Delta_2^2}{2}\int_s^{t'}f_{E_N}(x)dx} \label{eqIVTEFF1}
\ee
\be
T_{\rm eff}^2=2E_2(s) + \frac{4J}{K_2\Delta_2^2f_{E_N}(s)} \frac{\partial E_2(s)}{\partial s} - \e Q(s) + \bigl( \eps Q(s) - \eps^2\frac{2E_2(s)}{K_1K_2} \bigr) K_1\Delta_1^2\int_s^t dt'f_{E_N}(t')e^{\frac{K_2\Delta_2^2-K_1\Delta_1^2}{2}\int_s^{t'}f_{E_N}(x)dx} \label{eqIVTEFF2}
\ee

The expression for the total effective temperature is just:
\be
T_{\rm eff}^T= \frac{\frac{\partial C_1(t,s)}{\partial s}+\frac{\partial C_1(t,s)}{\partial s}}{G_1 (t,s)+G_2(t,s)} \label{eqIVTEFF3} 
\end{equation}
where the correlations are given by expressions (\ref{eqIVC1eps},\ref{eqIVC2eps}), and the responses are given by (\ref{eqIVG1eps},\ref{eqIVG2eps}).   

As in the case without coupling, the interesting dynamics is when the
temperature of the bath is zero. In this case, the energies and the
overlap decay to zero logarithmically which implies that $ f_{E_N} (t) $
vanishes like $ 1/t$. A careful evaluation of the integrals contributing
to the $\eps^2$ term shown in equations
(\ref{eqIVTEFF1},\ref{eqIVTEFF2}) reveals that they are a function of
$t/s$ which stays finite provided that ratio is finite. As we discussed
in the previous uncoupled or sequential case the effective temperatures
(\ref{eqIVTEFF1},\ref{eqIVTEFF2}) have full sense when we consider times
$t/s$ finite so no appreciable transfer of energy between the two systems has
still occurred.

It is clear from the asymptotic expressions for the energies and the
overlap that in the long-time limit ($ s\rightarrow \infty):$ \be \bigl(
Q(s) - \eps\frac{2E_1(s)}{K_1K_2} \bigr)\approx \bigl( Q(s) -
\eps\frac{2E_2(s)}{K_1K_2} \bigr) \approx \frac{1}{\log^2(s)} \ee While
the energies themselves decay as $ 1/\log(s)$ the relative difference
$(Q(s)/E_1(s) - \eps\frac{2}{K_1K_2})$ decays like $1/\log(s)$.  Up to
order $\e^2 $ we may write, in the limit $ s \longrightarrow \infty $
(with $\frac{t}{s}$ finite):

\be T_{\rm eff}^1=2E_1(s) - \e Q(s)~~~~~,~~~~~T_{\rm eff}^2=2E_2(s) -
\e Q(s), \ee because the asymptotic values of the $ E_1(s) $ and $
E_2(s) $ are the same the  effective temperatures
for the subsystems become identical in the long-time limit. Note that the case with dynamic coupling or
parallel dynamics is qualitatively different from the case without
dynamic coupling or sequential, because now all the degrees
of freedom are at the same effective temperature in the long-time
limit. Moreover, if we consider the global system it is easy to prove
that the total effective temperature defined in (\ref{T1}) is, in the
limit $ s\rightarrow \infty $ with $ \frac{t}{s}$ finite:

\be T_{\rm eff}^T=T_{\rm eff}^1=T_{\rm eff}^2=2E(s) - \e Q(s) \ee
where $ E=E_1=E_2 $ and $Q$ are given by (\ref{eqIVaa}). This is a
consequence of the fact that the energies of the two systems equalize
due to the dynamic coupling. Then, the whole system has the same
effective temperature and we can define an effective temperature for
the global system using FDT. The situation is the same as in
equilibrium systems. If we have two systems in equilibrium at a
certain temperature $T$, FDT not only holds for each subsystem but
also holds for the whole system bringing the temperature of the bath
$T$. At higher-orders in $\eps$ we expect that all terms with be
subleading for $t/s$ to be finite and asymptotically all three
temperatures coincide.

If we restrict to the case in
which the coupling constant vanishes, $\eps=0$, then the systems are still
coupled only through the dynamics and we obtain the same qualitatively
results: \be T_{\rm eff}^T=T_{\rm eff}^1=T_{\rm eff}^2=2E(s) \ee with $
E(s)\approx \frac{J}{8\log(s)}$. We conclude that the dynamic
coupling does not allow the presence of more than one effective
temperature in the whole system because even in the absence of explicit
coupling in the Hamiltonian, the dynamics itself makes the energies to
equalize in the long-time limit regime.

\section{CONCLUSIONS}

In this paper we have solved exactly the dynamics of two systems of
harmonic oscillators. We focused our attention on the concept of the
effective temperature defined through the FDR eq.(\ref{eqX}). The
effective temperature, a parameter defined by a relation of the
correlation and response functions, has been introduced in the context
of glass theory in order to understand the physics behind the dynamic
behavior of these out-off-equilibrium systems. In this paper we hope
to have clarified some aspects behind the physical meaning of this
effective temperature.

We have studied two types of couplings between the two subsystems of
oscillators, both in an aging state, finding that the way we couple them
is crucial for the validity of the zero-temperature law in the
off-equilibrium regime to hold.  The two cases we studied are the
dynamically uncoupled or sequential case and the dynamically coupled or
parallel case. In short, for the sequential case the coupling between
the variables of the two subsystems in the resulting dynamics arises
only through the Hamiltonian term $\eps Q$. For the parallel case, the
variables of the two subsystems are simultaneously updated leading to
further interaction between the two subsystems (on top of the $\eps Q$
coupling term in the energy).

We have discovered that for the dynamically uncoupled or sequential case
the two subsystems asymptotically reach different effective temperatures
which never equalize. So the whole system is divided in two parts, each
part characterized by its own effective temperature. The explanation for
this odd behavior lies behind the time dependence of the off-equilibrium
thermal conductivity which decays very quickly to allow for an
asymptotic equalization of the two effective temperatures.  This raises
the question whether different interacting degrees of freedom do
eventually reach the same effective temperature in the asymptotic
regime, condition tightly related to the validity of the zeroth law for
the off-equilibrium aging state. Our conclusion is that the zeroth law
is probably valid but hardly effective due to the very small
conductivity between the two subsystems in the aging state.  A
calculation of the thermal conductivity in this model will be shown
elsewhere \cite{GR2} and reveals that it decreases very quickly with
time, the heat transfer being unable to compensate for the difference of
the effective temperatures of the two subsystems.

For the dynamically coupled or parallel case, the two effective
temperatures equalize and the two subsystems are in a sort of thermal
equilibrium between them in the aging state. Consequently, the union
of the two subsystems has an effective temperature which coincides
with the temperature of each subsystem. In this case, the direct
coupling of the two subsystems through the parallel dynamics makes the
conductivity much larger than in the sequential case so in this case a
zero-th law for the aging state is effective and holds. In fact, these
results are also valid when we consider the particular case $ \eps = 0
$ in which the dynamics in itself is enough to equalize the effective
temperatures.

From these two type of couplings the first one is the only realistic.
Dynamics in real structural glasses involves short scale motions of
atoms and coupling between the different degrees of freedom occurs at
the level of the energy or Hamiltonian and never at the level of the
dynamics (at least in the classical regime). The results of this paper explain then why different degrees
of freedom in structural glasses can stay at different effective
temperatures forever. The off-equilibrium conductivity or heat
transfer between the different degrees of freedom is small enough for
the equalization of the effective temperatures associated to the
different degrees to never occur. This explains why when we touch a piece
of glass we feel it at the room temperature. The heat transfer coming
from the hotter non-thermalized degrees of freedom is extremely small.
Before finishing we must note one particular feature of our model. All the calculations were done at zero temperature where the energy vanishes asymptotically. The fact that the energy (and consequently the conductivity) of the system is exhausted in the asymptotic limit can lead to a pathological behavior not present in structural glasses at finite temperature. Nevertheless, the fact that the thermal conductivity vanishes much faster than the energy itself, suggests that the vanishing of the conductivity is not related to zero-temperature dynamics.

In the present calculation we have
focused on the interaction between two subsystems, both in the aging
state. When one of the subsystems is in an aging state and the other is
in equilibrium the analysis proceeds similarly, the conclusion being
that the non-thermalized subsystem determines the rate of heat transfer
and hence the measurement of the effective temperature. The value of the
effective temperature measured by a thermometer and other related
questions can be analyzed in detail in the present model and will be
presented elsewhere \cite{GR2}.

To conclude, although a zero-th law for non-equilibrium glassy systems
may hold, it is hardly effective because of the small energy transfer
occurring between degrees of freedom at different effective temperatures.
It would be very interesting to pursue this investigation further by
studying other solvable examples and showing that what we have
exemplified here is a generally valid for structural glasses as well as
for other glassy systems.

{\bf Acknowledgements.} We are grateful to M. Picco for a careful
reading of the manuscript. A. G. is supported by a grant from the
University of Barcelona. F. R is supported by the Ministerio de
Educaci\'on y Ciencia in Spain (PB97-0971).

\newpage
\begin{center}
 \textbf{APPENDIX A: A SHORT REVIEW OF THE BPR MODEL}
\end{center}
In this appendix we show how derive the results for the correlation and the response functions in ~\cite{BPR} in order
to understand the techniques we will use throughout this paper. In that
model, the system is constituted by $N$ uncoupled harmonic oscillators
which evolve with Monte Carlo dynamics. The energy of this system is

\begin{equation}
E({x_i}) = \frac{1}{2} K \sum_i x_i^2~~~~~.
\end{equation}

The result for the dynamical evolution for the energy is:
\begin{equation}
\frac{\partial E}{\partial t} = \frac{a_c}{2} \left(\frac{1- 4E\beta}{a_c\beta} f(t) + erfc(\alpha ) \right)~~~~~,\label{energy}
\end{equation}
where we have defined the quantities
\begin{equation}
a_c=\frac{1}{2} K\Delta^2~~~~~,~~~~~\alpha = \sqrt{\frac{K\Delta^2}{16E}},~~~~
erfc(x)=\frac{2}{\sqrt{\pi}} \int_x^\infty \exp(-x^2) dx,
\label{eqIIIerfc}
\end{equation}
\begin{equation}
f(t) = a_c\beta \exp \left(-\beta \frac{K\Delta^2}{2} (1-2E(t)\beta) \right) erfc(\alpha (t) (4E(t)\beta - 1))~~~~~.
\label{ft}
\end{equation}

The stationary solution is just $ E=\frac{1}{2}T$ as expected. Another important quantity is the acceptance rate which is the number of accepted Monte Carlo movements at a time $t$:
\begin{equation}
A(t)=\frac{1}{2} \left( \frac{f(t)}{a_c\beta} + erfc(\alpha ) \right)~~~~~.
\end{equation}

In the same way we can compute the equation for the correlation function defined as:
\begin{equation}
C(t,s) =\frac{1}{N} \sum_{i=1}^N x_i(t)  x_i(s)~~~~~, 
\end{equation}
and the evolution of the correlation function is given by
the equation

\begin{equation}
\frac{\partial C(t,s)}{\partial t}= -f(t) C(t,s)~~~~~,
\end{equation}
where the quantity $ f(t) $ has been previously defined in (\ref{ft}). The solution for the correlation function (which depends explicitly on two times) is:
\begin{equation}
C(t,s) = \frac{2E(s)}{K} \exp\left( -\int_{s}^t f(x) dx \right)~~,
\label{ApC} 
\end{equation}
where we have to add the initial condition $ C(s,s) = \frac{2E(s)}{K}. $  
In order to compute the equation for the response function defined by,
\begin{equation}
G(t,s) = \left( \frac{\delta M(t)}{\delta h(s)} \right)_{h=0}~~,
\end{equation} 
we have to consider the Hamiltonian perturbed by a small external field
\begin{equation}
 H = \frac{K}{2} \sum_{i=1}^N x_i^2 -h \sum_i x_i~~~~~.
\end{equation}

Then we compute the dynamical evolution for the magnetization, which in our model is defined by: $ M = \frac{1}{N} \sum_{i=1}^N x_i$ yielding:
\begin{equation}
\frac{\partial M(t)}{\partial t} = - \left( M(t) - \frac{h}{K} \right) f_{E'}(t) \Theta (t-s)~~~~~,
\label{mt}
\end{equation}
where we have defined a 'new' energy as:
\begin{equation}
E' = E -Mh + \frac{h^2}{2K}~~~~~.
\end{equation}
The quantity $ f_{E'}(t) $ is identically defined as in (\ref{ft}) but
with the new energy $E'$. Note that in the case in which $ h=0 $ the
magnetization will always be zero because of the initial condition we
consider, i.e $ M(0) = 0 $. Also, when we compute the
evolution for the response function, the first term in the right hand
side of the (\ref{mt}) is just the response. Then we have to analyze
carefully the second term (which is proportional to the external
field) in the right hand side of (\ref{mt}). First of all, we write
\begin{equation}
G(t,s)\approx \frac{\Delta M(t)}{h \Delta s}~~~~~,~~~~~\frac{\partial G(t,s)}{\partial t} \approx \frac{\frac{\Delta M(t)}{\Delta t}}{h \Delta s}~~~~~.
\end{equation}

We consider the variation of the magnetization as follows
\begin{equation}
\frac{\Delta M}{\Delta t} = \Delta s \frac{\partial}{\partial
s}\left(\frac{\partial M(t)}{\partial t}\right) \Theta (t-s) +  \Delta
s\frac{\partial M(t)}{\partial t} \delta (t-s)~~~~~,  \label{DeltaM} 
\end{equation}
and by keeping only the linear term in $ h $  in the second term of the
r.h.s in (\ref{DeltaM}) we get
\begin{equation}
\frac{\Delta M}{\Delta t} = \Delta s \frac{h}{K} \frac{\partial f(t)}{\partial s} \Theta (t-s) +  \Delta s \frac{h}{K} f(t) \delta (t-s)~~~~~.\label{sct}
\end{equation}

The first term in the r.h.s of (\ref{sct}) is obviously zero and only
the the second term gives a non-vanishing contribution. So the evolution for the
response function is
\begin{equation}
\frac{\partial G(t,s)}{\partial t} = -f(t) \left( G(t,s) - \frac{1}{K} \delta(t-s) \right)~~~~~,
\end{equation}
whose solution is
\begin{equation}
G(t,s) = \frac{f(s)}{K} \exp\left( -\int_{s}^t f(x) dx \right) \Theta (t-s)~~~~~.\label{ApG}
\end{equation}

Now, we are in position to compute the effective temperature based on the violation of FDT
\begin{equation}
T_{\rm eff} = \left( \frac{\frac{\partial C(t,s)}{\partial s}}{G (t,s)} \right) =2E(s) + \frac{1}{f(s)} \frac{\partial E(s)}{\partial s}~~~~~.
\end{equation}

Note that the effective temperature only depends on the smallest
time $s$. This feature is due to the simplicity of the model. In this model,
due to the finite amplitude of the Monte Carlo movements the system
never reaches the ground state $\lbrace x_i=0\rbrace$. In fact, Monte
Carlo dynamics induces entropic barriers which manifest as activated
behavior for the relaxation time. The interesting dynamics is found
when we study the relaxation of the system at zero temperature. To
obtain the dynamical equations at zero temperature we have to consider
only the negative changes in the energy. It can be seen that in the long time limit the relaxation of the energy is logarithmic
\begin{equation}
E(t) \approx \frac{K\Delta^2}{16} \frac{1}{ \log \left( \frac{2t}{\sqrt{\pi}} \right) + \frac{1}{2} \log(\log\left( \frac{2t}{\sqrt{\pi}} \right) )}\label{APAEN}~~~~~,
\end{equation}
moreover, we obtain the following asymptotic behavior of the function $f(t)$ and the acceptance rate:
\begin{equation}
f(t) \approx \frac{1}{t} \left( 1 + \frac{\frac{1}{2} \log(\log\left( \frac{2t}{\sqrt{\pi}} \right))}{\log \left(\frac{2t}{\sqrt{\pi}} \right)} \right)^{\frac{1}{2}},~~~
A(t) \approx \frac{1}{4t\log \left( \frac{2t}{\sqrt{\pi}} \right)}.
\end{equation}

For the long time behavior of the correlation and
the response functions we obtain to leading order in $\log(s)/\log(t)$

\begin{equation}
C(t,s)=\frac{2E(s)}{K} C_{norm}(t,s)~~~~,~~~~C_{norm}(t,s) \approx
\frac{s\log^2 \left( \frac{2s}{\sqrt{\pi}} \right)}{t\log^2 \left(
\frac{2t}{\sqrt{\pi}} \right)}~~~~,~~~~G(t,s) \approx \frac{1}{Kt}\left(\frac{ \log \left( \frac{2s}{\sqrt{\pi}} \right) }{\log\left( \frac{2t}{\sqrt{\pi}} \right)} \right)^2~~~~.
\end{equation}

\newpage
\begin{center}
\textbf{APPENDIX B: SOLUTION OF THE DYNAMICALLY UNCOUPLED OR SEQUENTIAL
CASE}
\end{center}

In this appendix we show explicitly the detailed calculations for the
case in which we sequentially update the two subsystems. Note that each
subsystem is updated in parallel but no simultaneous updating of the
whole system is performed so there is no direct coupling of the two
subsystems through the dynamics but only through an explicit coupling
term $\eps Q$ in the Hamiltonian. We have to take into account this fact
when we compute the distribution probability for a change in the
energy. The Hamiltonian we have to deal with is

\begin{equation} H = \frac{K_{1}}{2} \sum_{i=1}^N x_i^2 + \frac{K_2}{2} \sum_{i=1}^N y_i^2 - \epsilon \sum_{i=1}^N x_iy_i~~~~.\end{equation}

The main quantities we work with are

\begin{equation} E_1 = \frac{K_1}{2N} \sum_{i=1}^N x_i^2~~~~~,~~~~~E_2 =  \frac{K_2}{2N} \sum_{i=1}^N y_i^2~~~~~,~~~~~Q = \frac{1}{N}\sum_{i=1}^N x_i  y_i~~~~~, \label{qa} \end{equation} 
where $ E_1 $ and $ E_2 $ are the energy of the bare systems while $Q$ is the
overlap between them.

The Monte Carlo updating procedure is the following. First all the $
x_i $ are moved to $ x_i + r_i/\sqrt{N} $ where the $ r_i $ are random
variables Gaussian distributed with zero average and variance $
\Delta_1^2.$ The move is accepted according to a rule defined by an
acceptance probability $ W(\Delta E) $ which satisfies detailed
balance: $ W(\Delta E) = W(-\Delta E)\exp(-\beta \Delta E ), $ where $
\Delta E $ is the change in the Hamiltonian. Later all the $y_i$ are
moved to $ y_i+s_i/\sqrt{N} $, where the $ s_i $ are random variables
Gaussian distributed with zero average and variance $ \Delta_2^2. $
The same transition probability is now applied for the $y_i$
variables. This sequential updating of the $x_i$ and $y_i$ variables
is then iterated.  Note that the coupling in the dynamics only appears
through the change $\eps\Delta Q$ of the total energy. 

Now we compute the distribution probability of a change in the energy
of the first system. This probability distribution can be expressed

\be
P(\delta E_1) = \int_{-\infty}^\infty \delta \left( \delta E_1 - K_1\sum_i\left(\frac{r_ix_i}{\sqrt N} + \frac{r_i^2}{2N}\right) + \e \sum_i \frac{r_iy_i}{\sqrt{N}}\right) 
 \left( \prod_i \frac{dr_i}{\sqrt{2\pi\Delta_1^2}} \exp \left(-\frac{r_i^2}{2\Delta_1^2}\right)\right)
\end{equation}
and in the same way we can compute the probability for the other system:
\be
P(\delta E_2) = \int_{-\infty}^\infty \delta \left( \delta E_2 - K_2\sum_i\left(\frac{s_iy_i}{\sqrt N} + \frac{s_i^2}{2N}\right) + \e \sum_i \frac{s_ix_i}{\sqrt{N}}\right)
 \left( \prod_i \frac{ds_i}{\sqrt{2\pi\Delta_2^2}} \exp \left(-\frac{s_i^2}{2\Delta_2^2}\right)\right)~~~~~.
\end{equation}

Using the integral representation of the delta function:
\begin{equation}
\delta(m) =\frac{1}{2\pi} \int_{-\infty}^\infty \exp \left(i\lambda m\right)d\lambda~~~~~,\label{integral}
\end{equation}

we obtain

\begin{equation}
P(\delta E_1) = \frac{1}{\sqrt{4\pi K_1R_1\Delta_1^2}} \exp \left( - \frac{(\delta E_1 - \frac{K_1\Delta_1^2}{2})^2}{4K_1R_1\Delta_1^2} \right)~~~~~,
\end{equation}
\begin{equation}
P(\delta E_2) = \frac{1}{\sqrt{4\pi K_2R_2\Delta_2^2}} \exp \left( - \frac{(\delta E_2 - \frac{K_2\Delta_2^2}{2})^2}{4K_2R_2\Delta_2^2} \right)~~~~~,
\end{equation}
with the quantities
\begin{equation}
R_1 = E_1 -\epsilon Q + \frac{\epsilon^2 E_2}{K_1K_2}~~~~~,~~~~~R_2 = E_2 -\epsilon Q + \frac{\epsilon^2 E_1}{K_1K_2}~~~~~.
\end{equation}

Note that due to the explicit coupling $\epsilon$ the probability of a change in the energy of one system not only depends on their energy, but also on the energy of the other system and the overlap. Now, we can compute the evolution of the energies

\begin{equation}
\frac{\partial E_1}{\partial t} = \int_{-\infty}^0 d(\delta E_1) \delta E_1 P(\delta E_1) + \int_0^\infty d(\delta E_1) \delta E_1 P(\delta E_1) \exp(-\beta \delta E_1)~~~~~,
\end{equation}
\begin{equation}
\frac{\partial E_2}{\partial t} = \int_{-\infty}^0 d(\delta E_2) \delta E_2 P(\delta E_2) + \int_0^\infty d(\delta E_2) \delta E_2 P(\delta E_2) \exp(-\beta \delta E_2)~~~~~,
\end{equation}
yielding

\begin{equation}
\frac{\partial E_1}{\partial t} = - ( 2E_1 - \epsilon Q) f_{R_1}(t) + \frac{1}{2} \left( \frac{ f_{R_1}(t)}{\beta} + \frac{K_1\Delta_1^2}{2} ercf(\alpha_1) \right)\label{eqApBE1}~~~~~,
\end{equation}
\begin{equation}
\frac{\partial E_2}{\partial t} = - ( 2E_2 - \epsilon Q) f_{R_1}(t) + \frac{1}{2} \left( \frac{ f_{R_2}(t)}{\beta} + \frac{K_2\Delta_2^2}{2} ercf(\alpha_2) \right)\label{eqApBE2}~~~~~.
\end{equation}

We compute the equation for the evolution of the overlap in two steps. The first is the change in the overlap when the variables of the first system are moved; and the second one is when the variables of the second system are moved. So we must to compute two joint probability distributions

\be
P(\delta E_1,\delta Q) = \int_{-\infty}^\infty \delta \left( \delta E_1 - K_1\sum_i\left(\frac{r_ix_i}{\sqrt N} + \frac{r_i^2}{2N}\right) + \e \sum_i \frac{r_iy_i}{\sqrt{N}}\right)\delta \left( \delta Q - \sum_i \frac{r_iy_i}{\sqrt{N}} \right)
 \left( \prod_i \frac{dr_i}{\sqrt{2\pi\Delta_1^2}} \exp \left(-\frac{r_i^2}{2\Delta_1^2}\right)\right)
\end{equation}

\be
P(\delta E_2,\delta Q) = \int_{-\infty}^\infty \delta \left( \delta E_2 - K_2\sum_i\left(\frac{s_iy_i}{\sqrt N} + \frac{s_i^2}{2N}\right) + \e \sum_i \frac{s_ix_i}{\sqrt{N}}\right)\delta \left( \delta Q - \sum_i \frac{s_ix_i}{\sqrt{N}} \right)
 \left( \prod_i \frac{ds_i}{\sqrt{2\pi\Delta_2^2}} \exp \left(-\frac{s_i^2}{2\Delta_2^2}\right)\right)
\end{equation}

Then we compute the evolution equation for the overlap in each step and sum the two equations
\begin{equation}
\frac{\partial Q(1st))}{\partial t} = \int_{-\infty}^0 d(\delta E_1)\int_{-\infty}^\infty d(\delta Q)\delta Q P(\delta E_1,\delta Q)+\int_0^\infty d(\delta E_1) e^{-\beta \delta E_1}\int_{-\infty}^\infty d(\delta Q)\delta Q P(\delta E_1,\delta Q)~~~~~,     
\end{equation}

\begin{equation}
\frac{\partial Q(2nd)}{\partial t} = \int_{-\infty}^0 d(\delta E_2)\int_{-\infty}^\infty d(\delta Q)\delta Q P(\delta E_2,\delta Q)+\int_0^\infty d(\delta E_2) e^{-\beta \delta E_2}\int_{-\infty}^\infty d(\delta Q)\delta Q P(\delta E_2,\delta Q)~~~~~.     
\end{equation}

The solution of these equations is

\begin{equation}
\frac{\partial Q(1st)}{\partial t} = -\left( Q - \frac{2\epsilon E_2}{K_1K_2} \right) f_{R_1}(t)~~~~~,~~~~~\frac{\partial Q(2nd)}{\partial t} = -\left( Q - \frac{2\epsilon E_1}{K_1K_2} \right) f_{R_2}(t) 
\end{equation} 
which yields the final equation

\be
\frac{\partial Q}{\partial t} = -\left( Q - \frac{2\epsilon E_2}{K_1K_2}
\right) f_{R_1}(t)  -\left( Q - \frac{2\epsilon E_1}{K_1K_2} \right)
f_{R_2}(t)~~~\label{eqApBE3},
\ee

\noindent
with the quantities defined in
(\ref{eqIIIf},\ref{eqIIIerfc}). In the same way we can
compute the equation for the correlation and overlap functions defined
in (\ref{a},\ref{b}).  To compute their evolution
equations we must evaluate the joint probability of a change in the
energy and a change in the correlation function. Note that when we
consider the change in the variables $ x_i $ we have to consider the
energy of the system one, and when we consider the change in the
variables $ y_i $ we have to take into account the energy of the other
system. The joint probability can be decomposed into the probability
distribution for a change in the energy multiplied by a conditional
probability

\begin{equation}
P(\delta E_1,\delta C_1) = P(\delta E_1) P(\delta C_1 | \delta E_1)~~~~,~~~~
P(\delta E_2,\delta C_2) = P(\delta E_2) P(\delta C_2 | \delta E_2)~~~~,\ee
\be
P(\delta E_2,\delta Q_1) = P(\delta E_2) P(\delta Q_1 | \delta E_2)~~~~,~~~~
P(\delta E_1,\delta Q_2) = P(\delta E_1) P(\delta Q_2 | \delta E_1)~~~~.
\end{equation}

Then the evolution for the correlation functions can be computed using
\begin{equation}
\frac{\partial C_1(t,s)}{\partial t} = \int_{-\infty}^0 d(\delta E_1)\int_{-\infty}^\infty d(\delta C_1)\delta C_1 P(\delta E_1,\delta C_1)+\int_0^\infty d(\delta E_1) e^{-\beta \delta E_1}\int_{-\infty}^\infty d(\delta C_1)\delta C_1 P(\delta E_1,\delta C_1)~~~~~,     
\end{equation}

\begin{equation}
\frac{\partial C_2(t,s)}{\partial t} = \int_{-\infty}^0 d(\delta E_2)\int_{-\infty}^\infty d(\delta C_2)\delta C_2 P(\delta E_2,\delta C_2)+\int_0^\infty d(\delta E_2) e^{-\beta \delta E_2}\int_{-\infty}^\infty d(\delta C_2)\delta C_2 P(\delta E_2,\delta C_2)~~~~~,     
\end{equation}
\begin{equation}
\frac{\partial Q_1(t,s)}{\partial t} = \int_{-\infty}^0 d(\delta E_2)\int_{-\infty}^\infty d(\delta Q_1)\delta Q_1 P(\delta E_2,\delta Q_1)+\int_0^\infty d(\delta E_2) e^{-\beta \delta E_2}\int_{-\infty}^\infty d(\delta Q_1)\delta Q_1 P(\delta E_2,\delta Q_1)~~~~~,     
\end{equation}
\begin{equation}
\frac{\partial Q_2(t,s)}{\partial t} = \int_{-\infty}^0 d(\delta E_1)\int_{-\infty}^\infty d(\delta Q_2)\delta Q_2 P(\delta E_1,\delta Q_2)+\int_0^\infty d(\delta E_1) e^{-\beta \delta E_1}\int_{-\infty}^\infty d(\delta Q_2)\delta Q_2 P(\delta E_1,\delta Q_2)~~~~~,     
\end{equation}

yielding
\begin{equation}
\frac{\partial}{\partial t} \left( \begin{array}{c}
C_1(t,s) \\ C_2(t,s) \\ Q_1(t,s) \\ Q_2(t,s)
\end{array}\right) =
 -\left( \begin{array}{cccc}
f_{R_1}(t) & 0 & -\frac{\epsilon}{K_1}f_{R_1}(t) & 0 \\
0 & f_{R_2}(t) & 0 &  -\frac{\epsilon}{K_2}f_{R_2}(t) \\
 -\frac{\epsilon}{K_2}f_{R_2}(t) & 0 &  f_{R_2}(t) & 0 \\
0 &  -\frac{\epsilon}{K_1}f_{R_1}(t) & 0 & f_{R_1}(t) \end{array}\right)
\left( \begin{array}{c}
C_1(t,s) \\ C_2(t,s) \\ Q_1(t,s) \\ Q_2(t,s) \end{array} \right)~~~~~. 
\end{equation}

For the response functions we have to compute the dynamic evolution
equations for the magnetizations. We consider an external field coupled
to each system, so the new Hamiltonian is
\begin{equation}
 H = \frac{K_{1}}{2} \sum_{i=1}^N x_i^2 + \frac{K_2}{2} \sum_{i=1}^N y_i^2 - \sum_i(h_1x_i + h_2y_i) - \e \sum_i x_iy_i~~~~~.
\end{equation} 
We define the magnetizations as follows
\be
M_1=\sum_{i=1}^N x_i~~~~~,~~~~~M_2=\sum_{i=1}^N y_i~~~~~.
\end{equation}

Then, we perform the same steps as we did for the other
quantities. First of all we have to compute the joint probability of a
change in the magnetization and a change in the energy. For example, for
computing the response function for the first system we make $ h_2 = 0 $
and $ h_1 \neq 0 ; $ then we compute the joint probability distribution
for a change in $ M_1 $ and $ E_1 . $ After that we can obtain the
evolution for the magnetization of this system
\begin{equation}
\frac{\partial M_1(t)}{\partial t} = - \left( M_1(t) - \frac{h_1}{K_1} - \frac{\epsilon M_2(t)}{K_1} \right) f_{A_1}(t)\label{eqApM1}
\end{equation}   
\begin{equation}
A_1 = R_1 + \frac{h_1^2\Delta_1^2}{2} - h_1K_1\Delta_1^2 M_1 + \epsilon h_1\Delta_1^2 M_2\label{eqApA1}
\end{equation}
Note that in this case we are considering $ h_2 = 0$ but still the
equation for $M_1$ depends on $M_2$. For the sequential updating
procedure we have to consider the evolution for $ M_2(t)$ with
$ h_1 = 0 $ and $ h_2 \neq 0 $ which is, by symmetry considerations  
\begin{equation}
\frac{\partial M_2(t)}{\partial t} = -  \left( M_2(t) - \frac{h_2}{K_2} - \frac{\epsilon M_1(t)}{K_2} \right) f_{A_2}(t)~~~~~,\label{eqApM2}
\end{equation}
\begin{equation}
A_2 = R_2 + \frac{h_2^2\Delta_2^2}{2} - h_2K_2\Delta_2^2 M_2 + \epsilon h_2\Delta_2^2 M_1~~~~~.\label{eqApA2}
\end{equation}

We finally get the equations for the four different response functions
using the same procedure we
followed for the single system (see Appendix A). This yields

\beq                 
\frac{\partial G_1(t,s)}{\partial t}=- \left( G_1(t,s)f_{R_1}(t) - \frac{f_{R_1}(t)}{K_1} \delta(t-s) -\frac{\epsilon}{K_1} G_2'(t,s) f_{R_1}(t) \right)~~~~~,\\  
\frac{\partial G_2(t,s)}{\partial t}=- \left( G_2(t,s)f_{R_2}(t) - \frac{f_{R_2}(t)}{K_2} \delta(t-s) -\frac{\epsilon}{K_2} G_1'(t,s) f_{R_2}(t) \right)~~~~~,\\  
\frac{\partial G_1'(t,s)}{\partial t}=- \left( G_1'(t,s)f_{R_1}(t) -\frac{\epsilon}{K_1} G_2(t,s) f_{R_1}(t) \right)~~~~~,  
\\
\frac{\partial G_2'(t,s)}{\partial t}=- \left( G_2'(t,s)f_{R_2}(t)-\frac{\epsilon}{K_2} G_1(t,s) f_{R_2}(t) \right)~~~~~.  
\eeq

In order to compute the effective temperatures we shall use a perturbative expansion in terms of the coupling constant described in Appendix C.

\subsection{Equilibrium results}

In equilibrium the matrices for correlations and responses can be
exactly diagonalised. The results are

\beq
C_1(t,s) = \frac{2E_1(s)}{K_1} \left( \frac{\lambda_2 - a}{\lambda_2-\lambda_1} \exp(\lambda_1 (t-s)) -\frac{\lambda_1 - a}{\lambda_2-\lambda_1} \exp(\lambda_2 (t-s)) \right) \nonumber\\
 + \frac{b}{\lambda_2-\lambda_1}Q(s) \left( \exp(\lambda_2 (t-s))- \exp(\lambda_1 (t-s)) \right)~~~~~,\label{ApeqC1}
\eeq

\beq
Q_1(t,s) = Q(s) \left( \frac{\lambda_2 - a}{\lambda_2-\lambda_1} \exp(\lambda_2 (t-s)) -\frac{\lambda_1 - a}{\lambda_2-\lambda_1} \exp(\lambda_1 (t-s)) \right) \nonumber\\
 + \frac{2E_1(s)}{K_1}\frac{c(\lambda_2-a)}{(\lambda_2-\lambda_1)(\lambda_1 - d)} \left( \exp(\lambda_2 (t-s))- \exp(\lambda_1 (t-s)) \right)~~~~~,\label{ApeqQ1}
\eeq

\be
G_1(t,s) = \frac{f_{R_1}(s)}{K_1} \left( \frac{\lambda_2 - a}{\lambda_2-\lambda_1} \exp(\lambda_1 (t-s)) -\frac{\lambda_1 - a}{\lambda_2-\lambda_1} \exp(\lambda_2 (t-s)) \right)~~~~~, \label{ApeqG1}
\end{equation}

\noindent
with the values of the constants
\be
a=- f_{R_1}~~~~~,~~~~~b=\frac{\epsilon}{K_1}  f_{R_1}~~~~~,~~~~~c=\frac{\epsilon}{K_2}  f_{R_2}~~~~~,~~~~~d=- f_{R_2}~~~~~,
\end{equation}
\noindent
and the two eigenvalues:
\begin{eqnarray}
\lambda_1= \frac{a+d}{2}+\frac{\sqrt{(a+d)^2 - 4(ad-cb)}}{2}~~~~~,~~~~~~\lambda_2= \frac{a+d}{2}-\frac{\sqrt{(a+d)^2 - 4(ad-cb)}}{2}~~~~~.
\label{eigen2}
\end{eqnarray}

The results for the other two correlation functions have the same form

\beq
C_2(t,s) = \frac{2E_2(s)}{K_2} \left( \frac{\lambda_2 - a}{\lambda_2-\lambda_1} \exp(\lambda_1 (t-s)) -\frac{\lambda_1 - a}{\lambda_2-\lambda_1} \exp(\lambda_2 (t-s)) \right) \nonumber\\
 + \frac{b}{\lambda_2-\lambda_1}Q(s) \left( \exp(\lambda_2 (t-s))- \exp(\lambda_1 (t-s)) \right)~~~~~,\label{ApeqC2} 
\eeq

\beq
Q_2(t,s) = Q(s) \left( \frac{\lambda_2 - a}{\lambda_2-\lambda_1} \exp(\lambda_2 (t-s)) -\frac{\lambda_1 - a}{\lambda_2-\lambda_1} \exp(\lambda_1 (t-s)) \right) \nonumber\\
 + \frac{2E_2(s)}{K_2}\frac{c(\lambda_2-a)}{(\lambda_2-\lambda_1)(\lambda_1 - d)} \left( \exp(\lambda_2 (t-s))- \exp(\lambda_1 (t-s)) \right)~~~~~,\label{ApeqQ2}
\eeq

\be
G_2(t,s) = \frac{f_{R_2}(s)}{K_2} \left( \frac{\lambda_2 - a}{\lambda_2-\lambda_1} \exp(\lambda_1 (t-s)) -\frac{\lambda_1 - a}{\lambda_2-\lambda_1} \exp(\lambda_2 (t-s)) \right)~~~~~,\label{ApeqG2}
\end{equation}

\noindent
with the new values of the constants \be
a=-f_{R_2}~~~~~,~~~~~b=\frac{\epsilon}{K_2}
f_{R_2}~~~~~,~~~~~c=\frac{\epsilon}{K_1}
f_{R_1}~~~~~,~~~~~d=-f_{R_1}\label{finale}
\end{equation}
and the same expressions as in eqs.(\ref{eigen2}) for
$\lambda_1,\lambda_2$.

\newpage
\begin{center}
\textbf{APPENDIX C: SOLUTION FOR THE OFF-EQUILIBRIUM CORRELATIONS AND
RESPONSES IN THE INTERACTION REPRESENTATION}
\end{center}

In general we have to solve the following equation
\be
\frac{\partial \vec{v}}{\partial t} = A(t) \vec{v}~~~~~,
\end{equation}
with the initial condition $ \vec{v} (t)=\vec{v} (s) $. $A(t)$ is the
matrix with the time-dependent coefficients of our problem. It can be
decomposed as: \be A(t) = A_0(t) + \e A_I(t)~~~~~,
\end{equation}
where $ A_0(t) $ is the diagonal part and $ A_I(t)$ is the interaction
part of the matrix. We work in the interaction
representation. Therefore we start by doing the transformation \be
\vec{w}(t) = \left( \exp \left( -\int_s^t A_0(t')dt' \right) \right)
\vec{v}(t)~~~~~.
\end{equation}

The derivative of this new vector is simply:
\be
\frac{d\vec{w}}{dt}=\e~ \exp \left( -\int_s^t A_0(t')dt' \right) A_I(t)\vec{v}(t)~~~~~,
\end{equation}
which can be written as
\be
\frac{d\vec{w}}{dt}=\e B_I(t)\vec{w}(t)~~~~~,
\label{w}
\end{equation}
where
\be
B_I(t)= \exp \left( -\int_s^t A_0(t')dt' \right) A_I(t) \exp \left( \int_s^t A_0(t')dt' \right)~~~~~. 
\end{equation}

Now we must solve (\ref{w}) with the initial condition $  \vec{w}
(s)= \vec{v} (s)$. The formal solution for this equation is
\be
\vec{w}(t)=\vec{w}(s) + \e~\int_s^t B_I(t')\vec{w}(t')~~~~~,
\end{equation}
or equivalently

\be
\vec{w}(t)={\cal T}\exp\bigl(\eps\int_{s}^tB_I(s')\bigr)\vec{w}(s)~~~~~.
\ee

Where $ \cal T $ stands for the time ordered product. 
This equation can be iterated and solved to any order in $\eps$. Up to order $\eps^2$ we find
\be
Order~ zero:~~~\vec{w}(t)=\vec{v}(s)~~~~~,
\end{equation}
\be
Order~ \e:~~~\vec{w}(t)=\vec{v}(s)+  \e~\int_s^t B_I(t')\vec{v}(s)~~~~~,
\end{equation}
\be
Order~ \e^2:~~~\vec{w}(t)=\vec{v}(s)+ \e~\int_s^t B_I(t')\vec{v}(s) +\e^2~\int_s^t dt'\int_s^{t'}dt'' B_I(t') B_I(t'')\vec{v}(s)~~~~~.
\end{equation}

This is the procedure we have used in order to obtain the equations for
the responses and correlations for the dynamically coupled and uncoupled
cases.

\end{document}